\documentclass[aps,prc,twocolumn,superscriptaddress,showpacs]{revtex4}

\usepackage{graphicx}
\usepackage{dcolumn}
\usepackage{bm}
\usepackage{ulem}
\usepackage{color}

\newcommand{\al}{$\alpha$}

\newcommand{\raa}{($\alpha$,$\alpha$)}

\newcommand{\raapr}{($\alpha$,$\alpha'$)}
\newcommand{\rag}{($\alpha$,$\gamma$)}
\newcommand{\ran}{($\alpha$,n)}

\newcommand{\rap}{($\alpha$,p)}
\newcommand{\rgn}{($\gamma$,n)}

\newcommand{\rga}{($\gamma$,$\alpha$)}

\newcommand{\stot}{$\sigma_{\rm{reac}}$}
\newcommand{\sred}{$\sigma_{\rm{red}}$}
\newcommand{\ered}{$E_{\rm{red}}$}

\newcommand{\zniv}{$^{64}$Zn}

\newcommand{\gevii}{$^{67}$Ge}
\newcommand{\geviii}{$^{68}$Ge}
\newcommand{\gavii}{$^{67}$Ga}

\begin{document}

\title{
$\alpha$-scattering and $\alpha$-induced reaction cross sections of $^{64}$Zn
  at low energies
}
\author{A. Ornelas}
\affiliation{
Institute for Nuclear Research (MTA Atomki), H-4001 Debrecen, Hungary}
\author{P. Mohr}
\affiliation{
Institute for Nuclear Research (MTA Atomki), H-4001 Debrecen, Hungary}
\affiliation{
Diakonie-Klinikum, D-74523 Schw\"abisch Hall, Germany}
\author{Gy. Gy\"urky}
\email{gyurky@atomki.mta.hu}
\affiliation{
Institute for Nuclear Research (MTA Atomki), H-4001 Debrecen, Hungary}
\author{Z. Elekes}
\affiliation{
Institute for Nuclear Research (MTA Atomki), H-4001 Debrecen, Hungary}
\author{Zs. F\"ul\"op}
\affiliation{
Institute for Nuclear Research (MTA Atomki), H-4001 Debrecen, Hungary}
\author{Z. Hal\'asz}
\affiliation{
Institute for Nuclear Research (MTA Atomki), H-4001 Debrecen, Hungary}
\author{G.G. Kiss}
\altaffiliation[Present address:]{
RIKEN, 2-1 Hirosawa, Wako, Saitama 351-0198, Japan}
\affiliation{
Institute for Nuclear Research (MTA Atomki), H-4001 Debrecen, Hungary}
\author{E. Somorjai}
\affiliation{
Institute for Nuclear Research (MTA Atomki), H-4001 Debrecen, Hungary}
\author{T. Sz\"ucs}
\affiliation{
Institute for Nuclear Research (MTA Atomki), H-4001 Debrecen, Hungary}
\author{M. P. Tak\'acs}
\altaffiliation[Present address:]{
HZDR Dresden-Rossendorf, D-01314 Dresden, Germany}
\affiliation{
Institute for Nuclear Research (MTA Atomki), H-4001 Debrecen, Hungary}
\author{D. Galaviz}
\altaffiliation[Present address:]{
LIP Lisboa, 1000-149 Lisbon, Portugal}
\affiliation{
Centro de F\'{i}sica Nuclear, University of Lisbon, 1649-003 Lisbon, Portugal}
\author{R. T.  G\"uray}
\affiliation{
Kocaeli University, Department of Physics, TR-41380 Umuttepe, Kocaeli, Turkey}
\author{Z. Korkulu}
\affiliation{
Kocaeli University, Department of Physics, TR-41380 Umuttepe, Kocaeli, Turkey}
\author{N. \"Ozkan}
\affiliation{
Kocaeli University, Department of Physics, TR-41380 Umuttepe, Kocaeli, Turkey}
\author{C. Yal\c c\i n}
\affiliation{
Kocaeli University, Department of Physics, TR-41380 Umuttepe, Kocaeli, Turkey}
\date{\today}

\begin{abstract}
\begin{description}
\item[Background] $\alpha$-nucleus potentials play an essential role for the
  calculation of $\alpha$-induced reaction cross sections at low energies in
  the statistical model. Uncertainties of these calculations are related to
  ambiguities in the adjustment of the potential parameters to experimental
  elastic scattering angular distributions and to the energy dependence of the
  effective $\alpha$-nucleus potentials.
\item[Purpose] The present work studies the total reaction cross section
  $\sigma_{\rm{reac}}$ of $\alpha$-induced reactions at low energies which can
  be determined from the elastic scattering angular distribution or from the
  sum over the cross sections of all open non-elastic channels.
\item[Method] Elastic and inelastic $^{64}$Zn($\alpha$,$\alpha$)$^{64}$Zn
  angular distributions were measured at two energies around the Coulomb
  barrier at 12.1\,MeV and 16.1\,MeV. Reaction cross sections of the
  ($\alpha$,$\gamma$), ($\alpha$,$n$), and ($\alpha$,$p$) reactions were
  measured at the same energies using the activation technique. The
  contributions of missing non-elastic channels were estimated from statistical
  model calculations.
\item[Results] The total reaction cross sections from elastic scattering and
  from the sum of the cross sections over all open non-elastic channels agree
  well within the uncertainties. This finding confirms the consistency of the
  experimental data. At the higher
  energy of 16.1\,MeV, the predicted significant contribution of
  compound-inelastic scattering to the total reaction cross section is
  confirmed experimentally. As a by-product it is found that most recent
  global $\alpha$-nucleus potentials are able to describe the reaction cross
  sections for $^{64}$Zn around the Coulomb barrier.
\item[Conclusions] Total reaction cross sections of $\alpha$-induced
  reactions can be well determined from elastic scattering angular
  distributions. The present study proves experimentally that the total cross
  section from elastic scattering is identical to the sum of non-elastic
  reaction cross sections. Thus, the statistical model can reliably be used
  to distribute the total reaction cross section among the different open
  channels.
\end{description}
\end{abstract}

\maketitle

\section{Introduction}
\label{sec:intro}
The nucleosynthesis of so-called $p$-nuclei on the neutron-deficient side of
the chart of nuclides requires a huge reaction network including more than
1000 nuclei and more than 10000 nuclear reactions
(e.g.\ \cite{Woo78,Arn03,Rau06,Rap06}). The resulting production factors depend
sensitively on branchings between \rgn\ and \rga\ reactions which are located
several mass units ``north-west'' of stability on the chart of nuclides for
heavy nuclei and close to stability for nuclei in the mass range around $A
\approx 100$. It is impossible to measure all the required \rga\ reaction
rates in the laboratory. Instead, theoretical predictions have been used which
are based on statistical model (StM) calculations. In most cases it turns out
that the \rga\ reaction rate is better constrained by experimental \rag\ data
than by \rga\ data because of the strong influence of thermally excited states
in the target nucleus \cite{Mohr07,Rau11,Rau11b,Rau13}.

Calculations of \rga\ and \rag\ cross sections and reaction rates depend on
the \al\ transmission which in turn depends on the chosen \al -nucleus
potential. Angular distributions of \raa\ elastic scattering have been
measured with high precision in the last decade \cite{Mohr13,Pal12} to
determine the \al -nucleus potential at low energies. However, elastic
scattering at low energies is dominated by the Coulomb interaction, and the
angular distributions approach the Rutherford cross section for point-like
charges at astrophysically relevant energies. Consequently, \al -nucleus
potentials from elastic scattering are determined at somewhat higher energies
and have to be extrapolated down to the astrophysically relevant energies.

It has turned out over the years that standard \al -nucleus potentials like
the extremely simple McFadden-Satchler potential \cite{McF66} are able to
reproduce \ran\ and \rag\ cross sections around and above the Coulomb barrier,
whereas at very low energies below the Coulomb barrier (i.e., in the
astrophysically relevant energy range) an increasing overestimation of the
experimental reaction cross sections has been found in many cases (e.g.,
\cite{Som98,Gyu06,Ozk07,Cat08,Yal09,Gyu10,Kis11,Avr10,Sau11,Mohr11,Pal12b,Qui14,Qui15,Sch16}). Interestingly,
in the $A \approx 20 - 50$ mass range, experimental \rap\ and \ran\ data are
well described using the McFadden-Satchler potential \cite{Mohr15}. Several
alternative suggestions for low-energy \al -nucleus potentials have been made
in the last years in \cite{Su15,Avr15,Avr14,Mohr13,Sau11,Avr10,Dem02,Dem09},
and the related uncertainties of \al -induced reaction cross sections at low
energies were studied very recently by \cite{Per16,Avr16a,Mohr16a}.

The motivation of the present work is manyfold.  Firstly, we attempt to extend
the high-precision elastic scattering measurements of the last decade in the
mass range $89 \le A \le 144$ \cite{Mohr13} towards lower masses.  Secondly,
we include inelastic \raapr\ scattering into our analysis which may play a
significant role for the total (non-elastic) reaction cross section \stot\ at
very low energies \cite{Rau13b,Avr16}.  Thirdly, we have measured reaction
cross sections of the \ran , \rap , and \rag\ reactions at exactly the same
energy as \raa\ elastic and inelastic scattering. This avoids any
complications from the extrapolation of the energy-dependent \al -nucleus
potential. In our previous study \cite{Gyu12} such experimental data have been
used for a sensitive test of the basic quantum-mechanical equation which
relates the total reaction cross section \stot\ to the reflexion coefficients
$\eta_L$ of elastic scattering. In the present work 
the reduced experimental uncertainties
allow to use the quantum-mechanical relation for \stot\ to constrain the cross
sections of unobserved non-elastic channels. Note that the chosen target
nucleus \zniv\ is well-suited for such a study because most of the reaction
products of \al -induced reactions are unstable. This allows a simple and
robust determination of the total cross section for each reaction channel by
activation measurements.

The paper is organized as follows. In Sect.~\ref{sec:exp} we describe our
experimental procedures for the measurement of \al -induced reaction cross
sections on \zniv\ and \zniv \raa \zniv\ scattering. Sect.~\ref{sec:an_scat}
presents the analysis of our new scattering data and further scattering data
from literature. A comparison between the total reaction cross sections
\stot\ from scattering and from the sum of the individual reaction cross
sections will be given in Sect.~\ref{sec:comp}. The predictions of recent
global \al -nucleus potentials will be compared to our experimental reaction
data in Sec.~\ref{sec:pred}. A final discussion and conclusions will be
provided in Sect.~\ref{sec:summ}.

\section{Experimental set-up and procedure}
\label{sec:exp}
In order to give a comprehensive experimental description of the
\zniv\,+\,\al\ system, the cross sections of the following reaction channels
were measured in the present work: The elastic \al-scattering cross section at
E$_\alpha$\,=\,12.1 and 16.1 MeV was measured in a wide angular
range. Inelastic scattering leading to the first four exited states of
\zniv\ was also measured in an angular range limited to a more backward
region. Using the activation technique, the cross sections of the
\zniv\rag\geviii , \zniv\ran\gevii , and \zniv\rap\gavii\ reactions were also
measured at the same energies. Since the experimental techniques of both the
scattering and activation experiments were already described in detail
elsewhere \cite{Gyu12,Mohr13}, here only the most important features of the
measurements and the results are presented. 

\subsection{Scattering experiments}
\label{sec:exp_scat}

The scattering experiments were carried out at the cyclotron accelerator of
Atomki which provided \al\ beams of 12.05 and 16.12\,MeV energy with typical
intensity of 150\,nA. Targets were produced by evaporating highly enriched
(99.71\,\%) metallic \zniv\ onto thin (40\,$\mu$g/cm$^2$) carbon foils. The
thickness of the \zniv\ layer was about 150\,$\mu$g/cm$^2$ determined by
\al\ energy loss measurements. 
The energy loss at the energies of the scattering experiment is negligible
compared to the energy width of the beam.

\begin{figure}
\includegraphics[width=0.5\textwidth]{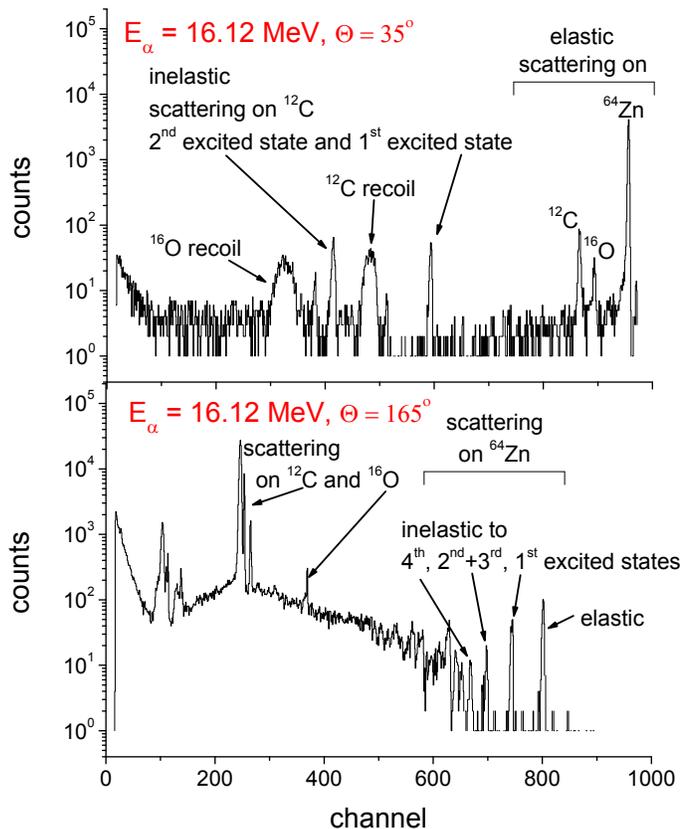}
\caption{\label{fig:scatteringspectra} Two typical scattering spectra measured at 16.12\,MeV \al-energy at forward (35$^\circ$, upper panel) and backward (165$^\circ$, lower panel) angles. The origin of the most prominent peaks are indicated.}
\end{figure}

The angular distributions were measured in a scattering chamber equipped with seven ion implanted Si particle detectors. The detectors were fixed on turntables enabling the measurement of scattered particle spectra in an angular range between 20$^\circ$ and 175$^\circ$. In addition, two detectors were fixed at $\pm$15$^\circ$ with respect to the beam. These monitor detectors were used for normalization purposes. Two typical scattering spectra are shown in Fig.\ref{fig:scatteringspectra} recorded in the forward and backward regions, respectively. Peaks corresponding to elastic and inelastic scattering events on \zniv\ and on carbon and oxygen target components are indicated.

The elastic scattering cross sections were measured at both energies in the
complete angular range between 20$^\circ$ and 175$^\circ$. The excitation
energies of the first four excited states of \zniv\ are 991.6\,keV ($2^+$),
1799.4\,keV ($2^+$), 1910.3\,keV ($0^+$), and 2306.8\,keV ($4^+$)
\cite{ENSDF,NDS}. The energy of the first excited state is far away enough
from both the ground state and the higher energy excited states so that the
peak corresponding to the inelastic scattering leading to this excited state
is well separated in the particle spectra. Therefore, the inelastic scattering
cross section for this excited state could be determined over a wider angular
range. The only limitation is caused by the elastic scattering events on
carbon and oxygen in the target which start to overlap with the inelastic peak
in the forward angle region. Therefore, the inelastic scattering cross section
to the first excited state could be only determined in the angular range between
40$^\circ$ and 175$^\circ$.

The second and third excited states could not be fully resolved because of their
energy difference of about 100\,keV. Therefore, only the sum
of the cross sections leading to these states could be measured in the angular
range between 60$^\circ$ and 175$^\circ$. Owing to the sufficient separation,
the inelastic cross section leading to the fourth excited state could be
determined from 60$^\circ$ to 175$^\circ$. Above the fourth exited state the
level density becomes too high, and hence no further inelastic cross sections
could be measured to higher lying levels. 

The experimental data will be provided to the community in tabular form
through the EXFOR database \cite{EXFOR}.

\subsection{Activation experiments}
\label{sec:exp_act}
The activation cross section measurements had been already carried out and
published in full details \cite{Gyu12}. In the present work only one
additional data point was determined since in \cite{Gyu12} no measurement was
carried out at 16.12\,MeV \al-energy, which is one of the energies where the scattering cross section was measured in the present work and the total cross section \stot\ is determined.

The new measurement at 16.12\,MeV \al-energy was carried out with exactly the
same conditions as the experiments of \cite{Gyu12}. The cyclotron accelerator
provided the \al-beam which bombarded a thin natural isotopic composition Zn
target on Al foil backing. The number of the \geviii , \gevii , and
\gavii\ isotopes produced by the \zniv\rag\geviii , \zniv\ran\gevii , and
\zniv\rap\gavii\ reactions, respectively, was determined off-line by measuring
the $\gamma$-radiation following the decay of the reaction products with a
shielded 100\,\% relative efficiency HPGe detector.

Table \ref{tab:actres} summarizes the results of the activation cross section
measurements. The first row at 12.05\,MeV \al-energy is taken from \cite{Gyu12}
while the second one at 16.12\,MeV is the result of the present work. The last
column shows the sum of the cross sections of the three reactions. These values
will be used in the next sections for the determination of the total cross
sections. 
\begin{table*}[htb]
\caption{\label{tab:actres} Cross section of the three measured reaction channels: The last column shows the sum of the cross sections of the three reactions where the calculation of the uncertainty takes into account the common systematic uncertainties. Details of the analysis can be found in \cite{Gyu12}, and the results in the first row are taken from \cite{Gyu12}.}
\begin{tabular}{cr@{\,$\pm$\,}lr@{\,$\pm$\,}lr@{\,$\pm$\,}lr@{\,$\pm$\,}lr@{\,$\pm$\,}l}
\hline
E$_\alpha$ & \multicolumn{2}{c}{E$^{\rm eff.}_{\rm c.m.}$} & \multicolumn{2}{c}{\zniv\rag\geviii\ } & \multicolumn{2}{c}{\zniv\ran\gevii} & \multicolumn{2}{c}{\zniv\rap\gavii\ } & \multicolumn{2}{c}{sum} \\
\cline{4-11}
(MeV) & \multicolumn{2}{c}{(MeV)} & \multicolumn{8}{c}{cross section (mbarn)} \\
\hline
12.05 & 11.29 & 0.09 & 2.42 & 0.21 & 109 & 12 & 255 & 28 & 366 & 40 \\
16.12 & 15.17 & 0.08 & 0.71 & 0.14 & 240 & 26 & 428 & 50 & 668 & 72 \\
\hline
\end{tabular}
\end{table*}

The energies $E_{\alpha,{\rm{lab}}}$ in the scattering experiments and in the
activation experiments were identical because exactly the same settings for
the cyclotron have been used. At the higher energy of 16.1\,MeV the effective
energies of the scattering and activation experiments are the same
($E_{\rm{c.m.}} = 15.17$\,MeV) because relatively thin targets were used
here. However, at the lower energy of 12.1\,MeV a thicker target had to be
used for the activation experiment leading to a slightly lower effective
energy in the activation experiment ($E_{\rm{c.m.}}^{\rm{eff}} = 11.29$\,MeV
compared to $E_{\rm{cm.}} = 11.34$\,MeV). The beamline for the scattering
experiments allows a more precise measurement of the beam energy using the
well-calibrated analyzing magnet in this beamline. Compared to the data in
\cite{Gyu12} ($E_{\rm{c.m.}}^{\rm{eff}} = 11.24 \pm 0.09$\,MeV), this leads to
a minor change of 50\,keV in the effective energy for the 12.1\,MeV activation
experiment. This minor change remains within the given uncertainties of
\cite{Gyu12}.

\section{Analysis of scattering data}
\label{sec:an_scat}
In a first step, the new elastic scattering data at low energies together with
data from literature are analyzed
in the optical model (OM). Fortunately, several angular distributions of \zniv
\raa \zniv\ elastic scattering at higher energies are also available in
literature. Therefore, we can study the energy dependence of the angular
distributions, the total reaction cross section \stot , and the resulting
optical model potentials (OMP). We restrict ourselves to data below about
50\,MeV.

The complex OMP $U(r)$ is given by
\begin{equation}
U(r) = V(r) + iW(r) + V_C(r)
\label{eq:OMP}
\end{equation}
with the real part $V(r)$ and the imaginary part $W(r)$ of the nuclear
potential and the Coulomb potential $V_C(r)$. The Coulomb potential is
calculated as usual from the model of a homogeneously charged sphere with the
Coulomb radius $R_C$. Various parametrizations have been used for the nuclear
potentials $V(r)$ and $W(r)$. 

For the imaginary part $W(r)$ Woods-Saxon (WS) potentials of volume and
surface type were applied:
\begin{equation}
W(r) = W_V f_V(x) + W_S \frac{df_S(x)}{dx}
\label{eq:WS}
\end{equation}
with the Woods-Saxon function
\begin{equation}
f_i(x) = [ 1 + \exp{(x)} ]^{-1}
\label{eq:WSf}
\end{equation}
where $x = (r-R_i A_T^{1/3})/a_i$ and $i = S,V$. $W_V$ and $W_S$ are the
strengths of the volume and surface imaginary potentials, $R_i$ are the radius
parameters, $a_i$ the diffuseness parameters, and $A_T^{1/3} = 4$ for the
target \zniv . Note that the maximum depth of a surface WS potential in the
chosen convention in Eq.~(\ref{eq:WSf}) is $W_S/4$. Following \cite{Mohr13},
at low energies below 25\,MeV only a surface imaginary potential was used in
combination with a real folding potential. At the higher energies a
combination of volume and surface Woods-Saxon potentials is necessary to
obtain excellent fits to the experimental angular distributions.

The real part of the nuclear potential was either taken as a volume WS
potential (similarly defined as above for the imaginary part) or calculated
from a folding procedure. Two parameters $\lambda$ for the strength and $w$
for the width were used to adjust the folding potential $V_F(r)$ to the
experimental data:
\begin{equation}
V(r) = \lambda \times V_F(r/w) \quad \quad .
\label{eq:fold}
\end{equation}
Obviously, the width parameter $w$ should remain close to unity; otherwise,
the folding procedure would become questionable. The strength parameter
$\lambda$ is typically around $1.1 - 1.4$ leading to volume integrals per
interacting nucleon pair $J_R \approx 320 - 350$\,MeV\,fm$^3$ for heavy nuclei
with a closed proton or neutron shell whereas slightly higher values have been
found for lighter nuclei and non-magic nuclei. Further details 
on the folding procedure and the chosen interaction can be found in
\cite{Mohr13}. 

The total (non-elastic) reaction cross section \stot\ is related to the
elastic scattering angular distribution by 
\begin{equation}
\sigma_{\rm{reac}} = \sum_L \sigma_L = 
   \frac{\pi}{k^2} \sum_L (2L+1) \, (1 - \eta_L^2) \quad \quad .
\label{eq:stot}
\end{equation}
Here $k = \sqrt{2 \mu E_{\rm{c.m.}}}/\hbar$ is the wave number,
$E_{\rm{c.m.}}$ is the energy in the center-of-mass (c.m.) system, and
$\eta_L$ are the real reflexion coefficients. These $\eta_L$ and the scattering
phase shifts $\delta_L$ define the angular distribution
$\left(\frac{d\sigma}{d\Omega}\right)(\vartheta)$ of elastic
scattering, whereas \stot\ depends only on the $\eta_L$, but is independent of
the $\delta_L$. The $\sigma_L$ are the contributions of the $L$-th partial
wave to the total reaction cross section \stot\ which show a characteristic
behavior as discussed e.g.\ in \cite{Mohr11,Mohr13b}.

\subsection{Elastic scattering data from literature}
\label{sec:lit}
\begin{figure}[htb]
\includegraphics[width=0.95\columnwidth,clip=]{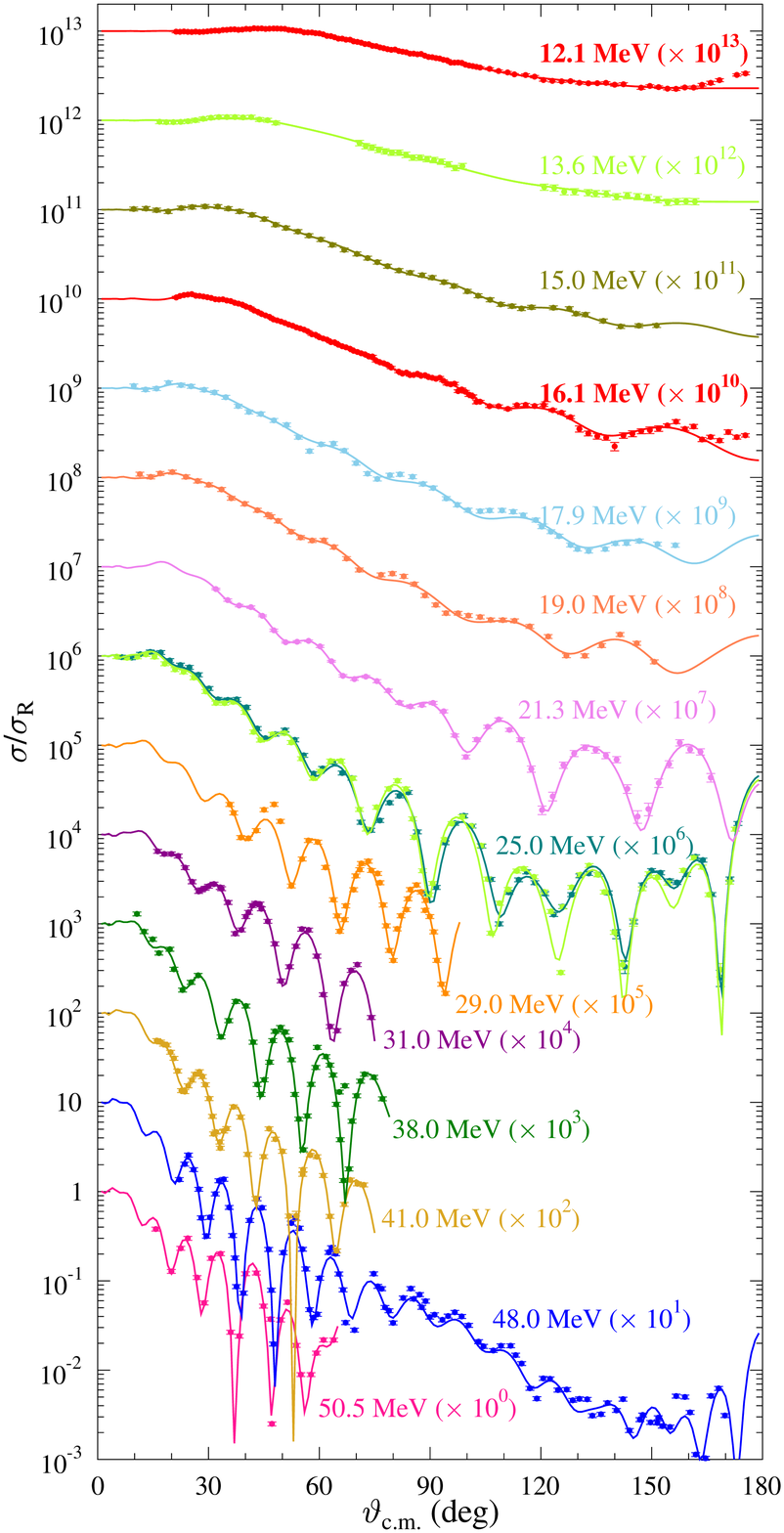}
\caption{
\label{fig:elast_all}
(Color online)
Elastic \zniv \raa \zniv\ scattering cross section at energies from 12 to
50\,MeV. The new data at 12.1 and 16.1\,MeV are shown in red and labeled in
bold. The other experimental data have been taken from DiPietro {\it et
  al.}\ \cite{DiP04} (13.6\,MeV), Robinson and Edwards \cite{Rob78} (15.0,
17.9, and 19.0\,MeV), Fulmer {\it et al.}\ \cite{Ful68} (21.3\,MeV), England
{\it et al.}\ \cite{Eng82} and Ballester {\it et al.}\ \cite{Bal88} (25\,MeV), 
Baktybaev {\it et al.}\ \cite{Bak75} (29, 38, and 50.5\,MeV), Alpert {\it et
  al.}\ \cite{Alp70} (31\,MeV), McDaniels {\it et al.}\ \cite{McD60}
(41\,MeV), and Pirart {\it et al.}\ \cite{Pir78} (48\,MeV). 
}
\end{figure}

Elastic \zniv \raa \zniv\ scattering has been studied in many experiments over
a broad energy range. Here we focus on the data up to energies of about
50\,MeV. The determination of an optical potential from angular distributions
requires high-quality scattering data over the full angular range. A careful
determination of the uncertainties is also mandatory because these
uncertainties have dramatic impact on the $\chi^2$ minimization
procedure. Therefore, we briefly review the status of the data from
literature as well as the availability and reliability of the data in the
EXFOR data base \cite{EXFOR}. The data are compared to the theoretical angular
distributions in Fig.~\ref{fig:elast_all}. The obtained OMP parameters are
listed in Table \ref{tab:pot}.

DiPietro {\it et al.}\ \cite{DiP04} have primarily studied \zniv
($^6$He,$^6$He)\zniv\ elastic scattering, and for comparison also the \zniv
\raa \zniv\ reaction was studied at $E_{\rm{lab}} = 13.2$\,MeV. The data cover
a broad angular range but there are also two larger gaps around $\vartheta
\approx 60^\circ$ and $110^\circ$. The data (including uncertainties) are
available from the EXFOR data base \cite{EXFOR}.

Robinson and Edwards \cite{Rob78} have measured angular distributions at
$E_{\rm{lab}} = 14.99$, 17.94, and 18.99\,MeV. The data cover a broad angular
range from very forward ($\vartheta \approx 10^\circ$) to backward angles
around $\vartheta \approx 150^\circ$. The data are not yet available at
EXFOR, and therefore the data had to be read from Fig.~2 in \cite{Rob78}. The
digitization has been performed using a high-resolution medical
scanner. However, the quality of the data is essentially limited by the
presentation of Fig.~2 in \cite{Rob78}. Error bars are typically smaller than
the symbols in Fig.~2 of \cite{Rob78}; therefore, a fixed uncertainty of 5\,\%
has been assigned to all data points. Note that the chosen absolute value of
5\,\% does affect the resulting $\chi^2/F$ but does not affect the $\chi^2$
minimization procedure and the resulting OMP.

The data of Fulmer {\it et al.}\ \cite{Ful68} at $E_{\rm{lab}} = 21.3$\,MeV
are available in tabular form (Table V of \cite{Ful68}). The data reach the
backward angular range up to $\vartheta \approx 165^\circ$. However, the data
unfortunately do not cover the forward region ($\vartheta \ge 32^\circ$), and
thus the absolute normalization cannot be fixed in the usual way by Rutherford
scattering. Here it is interesting to note that a much better description of
the experimental data can be obtained as soon as the absolute normalization of
the data is considered as a free parameter. The best fit is obtained using a
normalization of 0.787 for the data in Table V of \cite{Ful68}. Nevertheless,
the resulting parameters of the fits do not change dramatically; i.e., already
the shape of the angular distribution is able to constrain the fit reasonably
well. For completeness it should also be mentioned that the data shown in
Fig.~4 of \cite{Ful68} are also about 10\,\% lower than the values given in
Table V of \cite{Ful68}.

There are two data sets at $E_{\rm{lab}} = 25$\,MeV in EXFOR which reference
England {\it et al.}\ \cite{Eng82} and Ballester {\it et
  al.}\ \cite{Bal88}. Both data sets have been digitized from the given
Figures in \cite{Eng82,Bal88} (Fig.~2 of \cite{Eng82} and Fig.~3 of
\cite{Bal88}). The focus of \cite{Bal88} is inelastic scattering, and it is
explicitly stated that ``The elastic scattering data have already been
published$^{30}$.'' where the syperscript ``$^{30}$'' is a reference to
\cite{Eng82}. Thus, both data sets should be identical. However, due to
uncertainties of the digitization process, in fact both data sets show minor
discrepancies (and not even the number of data points agree). We have decided
to analyze both data sets; such an analysis can provide some insight into the
uncertainties of the resulting OMP parameters which result from the
re-digitization procedure. The angular distribution published in
\cite{Eng82,Bal88} covers the full angular range from forward angles
($\vartheta \approx 10^\circ$) to backward angles ($\vartheta \approx
170^\circ$) with small uncertainties (typically smaller than the shown point
size). Unfortunately, these small uncertainties are not available anymore, and
a fixed uncertainty of 5\,\% has been used in the fitting procedure. It turns
out that the resulting OMP parameters from the two EXFOR data sets are close to
each other. As with the analysis of the 21.3\,MeV data (see discussion below),
small discrepancies in the shape of the imaginary potential at large radii ($R
\gtrsim 8$\,fm) lead to noticable effects in the total reaction cross section
\stot\ which changes by about 10\,\% from 1317\,mb from the data of
\cite{Bal88} to 1481\,mb from the data of \cite{Eng82}. Therefore, we
recommend the average \stot\ = $1399 \pm 82$\,mb at 25\,MeV, and we assign the
same 6\,\% uncertainty to all \stot\ data which are derived from digitized
angular distributions.

The angular distributions at $E_{\rm{lab}} = 29$, 38, and 50.5\,MeV measured
by Baktybaev {\it et al.}\ \cite{Bak75} cover only a limited angular range
up to about $\vartheta \approx 90^\circ$. The data are available at EXFOR, but
have larger uncertainties of about $5 - 10$\,\% which are further
increased by the re-digitization. Consequently, potentials derived from these
data are not as reliable as in the other cases of this work. Further data from
the same institute \cite{Ais89} are available at EXFOR. But these data
\cite{Ais89} contain only very few points and are thus not included in the
present analysis.

The data at 31\,MeV by Alpert {\it et al.}\ \cite{Alp70} also cover only a
limited angular range. The above statements on the Baktybaev {\it et
  al.}\ data also hold here. As the data are not available at EXFOR, we have
re-digitized the angular distribution from Fig.~3 of \cite{Alp70}. Error bars
are not visible in this Fig.~3. We have used a fixed 5\,\% uncertainty in the
fitting procedure.

A limited angular distribution is available by McDaniels {\it et
  al.}\ \cite{McD60} at 41\,MeV. Re-digitized data without uncertainties are 
available at EXFOR. Again we have used a fixed 5\,\% uncertainty for the
fitting procedure of these data.

The data at 43\,MeV by Broek {\it et al.}\ \cite{Bro62} cover a very limited
angular range from about $10^\circ - 50^\circ$. In addition, the scanned pages
of this article (as provided at the ScienceDirect web page) are distorted. We
decided to exclude these data from our analysis.

An excellent angular distribution at 48\,MeV from about $20^\circ$ to
$170^\circ$ is available in Pirart {\it et al.}\ \cite{Pir78}. Unfortunately,
again the data in EXFOR had to be re-digitized from Fig.~1 of \cite{Pir78}
where uncertainties are not visible. Similar to the previous cases, we have
used a fixed 5\,\% uncertainty in our analysis.

For all angular distributions fits were performed in the following way. First,
a folding potential was calculated at the average energy of 21.3\,MeV using the
energy-dependent parameters of the nucleon-nucleon interaction listed in
\cite{Mohr13}. The energy dependence of these parameters is relatively weak
and has practically no impact on the final results. For a detailed discussion,
see \cite{Mohr13}. Next, the strength parameter $\lambda$ and the width
parameter $w$ of the folding potential and the parameters $W_i$, $R_i$, and
$a_i$ of the imaginary part were fitted simultaneously to the experimental
angular distributions. In addition, the absolute normalization $N$ of the
angular distributions was allowed to vary because this absolute normalization
often has much larger uncertainties. The normalization factors $N$ deviate
from unity by not more than about 30\,\%. Fits with fixed $N = 1.0$ lead in
many cases to much poorer $\chi^2$ of the fit. E.g., in the case of the
21.3\,MeV data of Fulmer {\it et al.}\ \cite{Ful68} $\chi^2$ reduces by about
a factor of 4 from fixed $N = 1.0$ to fitted $N = 0.787$. The resulting OMP
parameters remain relatively stable with variations of about 2\,\% for the
volume intgrals $J_R$ and $J_I$. However, relatively small changes in the
shape of the imaginary potential at large radii ($R \gtrsim 8$\,fm) result in
a change of the total reaction cross section \stot\ of about 10\,\% from
1200\,mb for fixed $N = 1.0$ to 1327\,mb for fitted $N = 0.787$. In cases
where the additional free parameter $N$ did not improve the reduced
$\chi^2/F$, a fixed normalization $N=1.0$ was used. The results of these fits
are shown in Fig.~\ref{fig:elast_all} and listed in Table \ref{tab:pot}. In
general, an excellent reproduction of the experimental angular distributions
could be achieved in the full energy range under study. Further information on
the uncertainties of the total reaction cross section \stot\ from elastic
scattering angular distributions is given in \cite{Mohr10}.
\begin{table*}[tbh]
\caption{\label{tab:pot}
Parameters of the optical potential and the total reaction
cross section \stot\ derived from \zniv \raa \zniv\ angular
distributions in literature
\cite{DiP04,Rob78,Ful68,Eng82,Bal88,Bak75,Alp70,McD60,Pir78} and from this
work (last two lines).
}
\begin{center}
\begin{tabular}{cccccccccrcccccr@{$\pm$}lc}
\multicolumn{1}{c}{$E_{\rm{lab}}$} 
& \multicolumn{1}{c}{$E_{\rm{c.m.}}$}
& \multicolumn{1}{c}{$\lambda$}
& \multicolumn{1}{c}{$w$}
& \multicolumn{1}{c}{$J_R$}
& \multicolumn{1}{c}{$r_{R,{\rm{rms}}}$}
& \multicolumn{1}{c}{$W_V$}
& \multicolumn{1}{c}{$R_V$}
& \multicolumn{1}{c}{$a_V$}
& \multicolumn{1}{c}{$W_S$}
& \multicolumn{1}{c}{$R_S$}
& \multicolumn{1}{c}{$a_S$}
& \multicolumn{1}{c}{$J_I$}
& \multicolumn{1}{c}{$r_{I,{\rm{rms}}}$}
& \multicolumn{1}{c}{$N$}
& \multicolumn{2}{c}{\stot\ \footnote{from the local potential fit using
    Eq.~(\ref{eq:stot}); uncertainties estimated as discussed in the text}}
& \multicolumn{1}{c}{Ref.} \\
\multicolumn{1}{c}{(MeV)} 
& \multicolumn{1}{c}{(MeV)}
& \multicolumn{1}{c}{(--)}
& \multicolumn{1}{c}{(--)}
& \multicolumn{1}{c}{(MeV\,fm$^3$)}
& \multicolumn{1}{c}{(fm)}
& \multicolumn{1}{c}{(MeV)}
& \multicolumn{1}{c}{(fm)}
& \multicolumn{1}{c}{(fm)}
& \multicolumn{1}{c}{(MeV)}
& \multicolumn{1}{c}{(fm)}
& \multicolumn{1}{c}{(fm)}
& \multicolumn{1}{c}{(MeV\,fm$^3$)}
& \multicolumn{1}{c}{(fm)}
& \multicolumn{1}{c}{(--)}
& \multicolumn{2}{c}{(mb)} 
& \multicolumn{1}{c}{Exp.} \\
\hline
13.4 & 12.40 & 1.314 & 1.019  & 377.7 & 4.757 
& \multicolumn{3}{c}{--} & 139.0 & 1.603 & 0.380 & 107.9 & 6.593 
& 0.946  &  610 & 80\footnote{uncertainty from \cite{Gyu12}} & \cite{DiP04} \\ 
15.0 & 14.11 & 1.292 & 1.029  & 378.3 & 4.808 
& \multicolumn{3}{c}{--} & 105.3 & 1.607 & 0.410 &  88.7 & 6.638 
& 1.0\footnote{fixed normalization $N$} &  858 &  52 & \cite{Rob78} \\ 
17.9 & 16.88 & 1.336 & 1.012  & 371.3 & 4.726 
& \multicolumn{3}{c}{--} & 102.5 & 1.530 & 0.468 &  89.8 & 6.404 
& 1.0\footnotemark[3] & 1083 &  65 & \cite{Rob78} \\ 
19.0 & 17.87 & 1.344 & 1.015  & 377.9 & 4.744 
& \multicolumn{3}{c}{--} & 111.0 & 1.525 & 0.456 &  94.1 & 6.371 
& 1.0\footnotemark[3] & 1149 &  69 & \cite{Rob78} \\ 
21.3 & 20.05 & 1.419 & 0.983  & 366.9 & 4.592 
&  $-24.0$  & 1.233 & 0.895   & $-15.4$ & 1.544 & 0.325 &  52.9 & 4.809 
& 0.787 & 1327 &  80 & \cite{Ful68} \\ 
%
25.0 & 23.53 & 1.435 & 0.980  & 367.6 & 4.578 
&  $-32.9$  & 1.080 & 0.985   & $-28.7$ & 1.544 & 0.255 &  54.9 & 4.659 
& 1.0\footnotemark[3] & 1481 &  89 & \cite{Eng82} \\ 
%
25.0 & 23.53 & 1.457 & 0.986  & 379.9 & 4.605 
&  $-22.2$  & 1.591 & 0.437   & $-79.0$ & 1.478 & 0.302 &  59.6 & 4.565 
& 1.0\footnotemark[3] & 1317 &  79 & \cite{Bal88} \\ 
29.0 & 27.29 & 1.386 & 0.998  & 374.2 & 4.659 
&   $-2.5$  & 1.963 & 0.338   & 214.3   & 1.245 & 0.336 & 109.0 & 5.369 
& 1.344 & 1458 &  88 & \cite{Bak75} \\ 
31.0 & 29.17 & 1.353 & 0.995  & 362.3 & 4.646 
&   $-3.9$  & 1.851 & 0.691   & 273.1   & 1.291 & 0.191 &  96.8 & 5.551 
& 0.894 & 1608 &  97 & \cite{Alp70} \\ 
38.0 & 35.76 & 1.419 & 1.010  & 397.7 & 4.717 
&  $-10.6$  & 1.816 & 0.339   & 322.4   & 1.318 & 0.062 &  95.1 & 5.628 
& 0.920 & 1580 &  95 & \cite{Bak75} \\ 
41.0 & 38.58 & 1.384 & 0.972  & 345.2 & 4.537 
&   $-1.8$  & 2.156 & 0.662   & 209.2   & 1.242 & 0.352 & 110.5 & 5.568 
& 0.768 & 1797 & 108 & \cite{McD60} \\ 
48.0 & 45.17 & 1.472 & 0.989  & 381.7 & 4.618 
&  $-28.5$  & 1.532 & 0.603   & $-16.6$ & 1.412 & 0.374 & 107.5 & 5.189 
& 1.135 & 1691 & 102 & \cite{Pir78} \\ 
50.5 & 47.52 & 1.375 & 0.990  & 363.1 & 4.624 
&   $-4.3$  & 1.961 & 0.419   & 204.8   & 1.267 & 0.178 & 80.9  & 5.640 
& 1.608 & 1721 & 103 & \cite{Bak75} \\ 
\hline
12.1 & 11.34 & 1.288 & 1.038  & 394.5 & 4.845 
& \multicolumn{3}{c}{--} & 186.5 & 1.697 & 0.275 & 116.6 & 6.876 
& 1.0\footnotemark[3]  
& 428 & 7\footnote{discussion of uncertainty: see text} 
& \footnote{this work} \\ 
16.1 & 15.17 & 1.330 & 1.006  & 371.2 & 4.696 
& \multicolumn{3}{c}{--} &  93.7 & 1.542 & 0.457 &  81.3 & 6.439 
& 1.0\footnotemark[3]  
& 905 & 18\footnotemark[4] 
& \footnotemark[5] \\ 
\hline
\end{tabular}
\end{center}
\end{table*}

\subsection{New elastic scattering data at 12 and 16\,MeV}
\label{sec:elast_new}
After the successful description of the elastic \zniv \raa \zniv\ scattering
data from literature 
\cite{DiP04,Rob78,Ful68,Eng82,Bal88,Bak75,Alp70,McD60,Pir78} we expected a
similar behavior for the analysis of our new data at 12.1 and
16.1\,MeV. However, 
the new data cover backward angles up to about $\vartheta \approx 175^\circ$
which exceeds the angular range of the literature data at low energies
\cite{DiP04,Rob78}. We found an unexpected increase of the
Rutherford-normalized cross section which is more pronounced at the lower
energy, and as it turns out it is practically impossible to describe the
full angular distributions using an OMP composed of a real folding potential
and an imaginary surface WS potential (similar to the fits to literature data
at low energies). The reduced $\chi^2/F$ of the OM fits does not reach values
around 1.0, but remains at about 2.0 (2.4) for the 12\,MeV (16\,MeV) data.

\begin{figure}[htb]
\includegraphics[width=0.95\columnwidth,clip=]{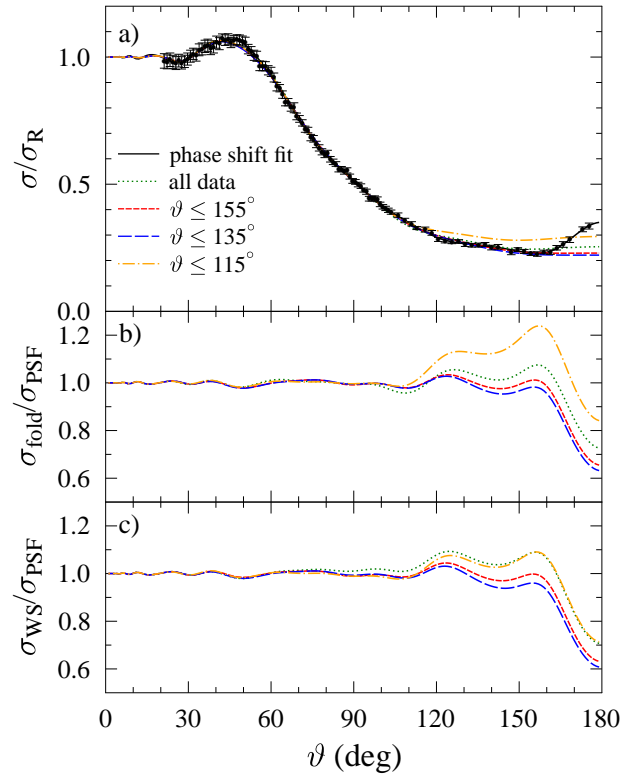}
\caption{
\label{fig:e12elast}
(Color online)
a) Elastic \zniv \raa \zniv\ scattering cross section at $E = 12.1$\,MeV,
compared to a phase shift analysis (PSF; full black line) and to OM fits using
a real folding and an imaginary surface Woods-Saxon potential. The fits use
the full data set (green dotted) and restricted data sets with 
$\vartheta \lesssim 155^\circ$ (red dashed), 
$\vartheta \lesssim 135^\circ$ (blue long-dashed), and
$\vartheta \lesssim 115^\circ$ (orange dash-dotted). 
The middle part b) shows these OM fits normalized to the PSF, and in the lower
part c) OM fits using volume Wood-Saxon potentials are again normalized to the
PSF. Further discussion see text. 
}
\end{figure}

\begin{figure}[htb]
\includegraphics[width=0.95\columnwidth,clip=]{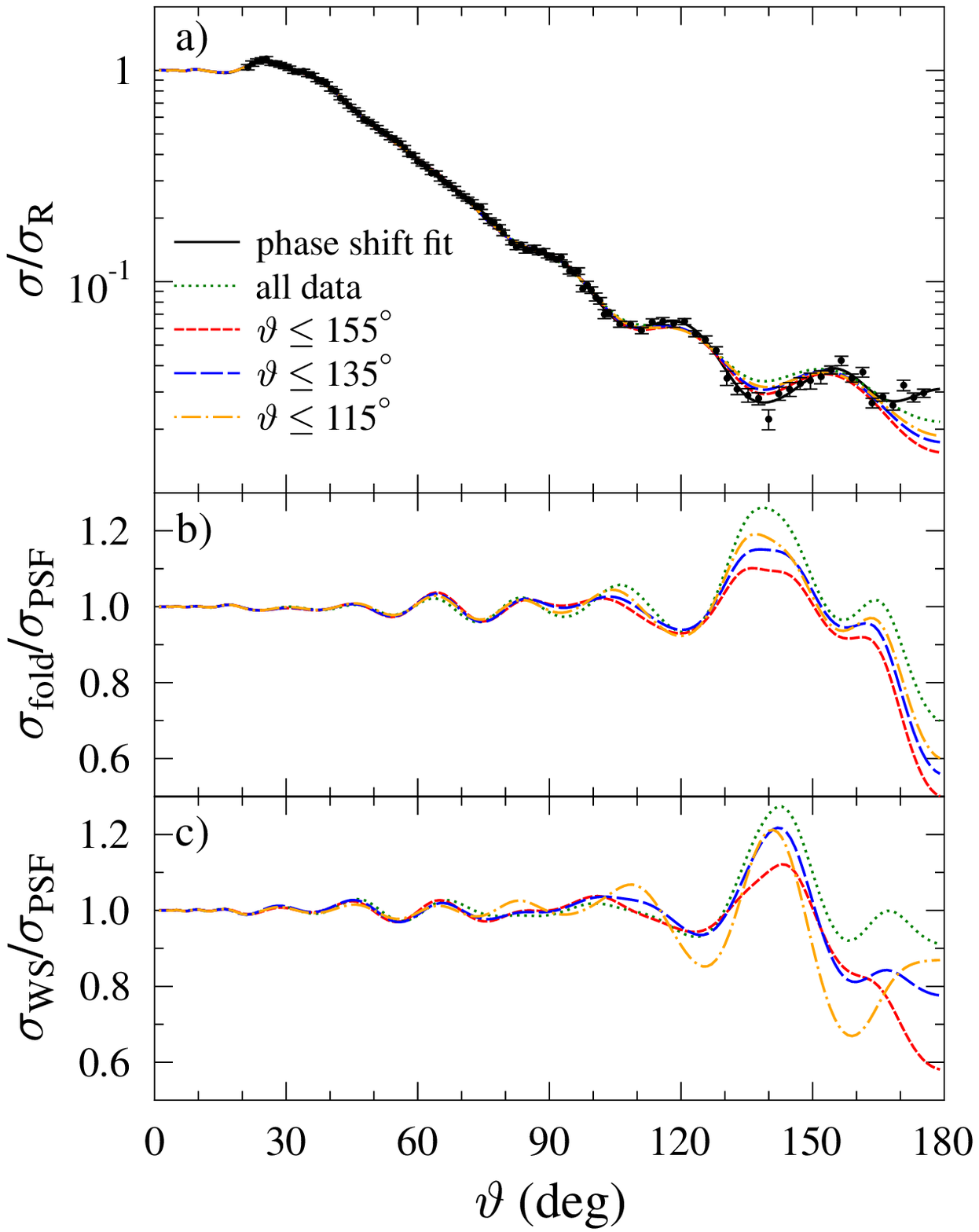}
\caption{
\label{fig:e16elast}
(Color online)
Same as Fig.~\ref{fig:e12elast}, but for the angular distribution at $E =
16.1$\,MeV.
}
\end{figure}

Therefore, we have performed a phase shift fit (PSF) using the method of
\cite{Chi96}. The PSFs are able to reproduce the full angular distributions at
both energies with $\chi^2/F \lesssim 1$ (see Figs.~\ref{fig:e12elast} and
\ref{fig:e16elast}). Values of $\chi^2/F \gg 1$ in a PSF would have been an
indication for experimental problems.

The comparison between the OM fits and the PSFs shows clearly that the poor
description of the angular distributions in the OM fits is related to the rise
of the elastic scattering cross section at extreme backward angles. Therefore,
we have made additional fits which have been restricted to data at $\vartheta
\lesssim 155^\circ$ ($135^\circ$, $115^\circ$). Already the restriction to
$\vartheta \lesssim 155^\circ$ leads to a dramatic improvement of $\chi^2/F$
at both energies by about a factor of two whereas the $\chi^2/F$ do not
improve further for the fits restricted to $135^\circ$ or
$115^\circ$. Therefore, we list the results for the OM fits truncated at
$\vartheta = 155^\circ$ in Table \ref{tab:pot}.

Although the $\chi^2/F$ of the various fits differ significantly, it is
difficult to visualize the differences. In the standard presentation (upper
parts of Figs.~\ref{fig:e12elast} and \ref{fig:e16elast}) the various fits are
mostly hidden behind the experimental data. Therefore, we consider the PSFs
(with $\chi^2/F \lesssim 1$) as quasi-experimental data and show the ratio
between the OMP fits and the PSFs in the middle part of
Figs.~\ref{fig:e12elast} and \ref{fig:e16elast}. Here it is nicely visible
that all truncated fits underestimate the most backward cross sections by
about $30-40$\,\%. As soon as the full data set is used for fitting, the
underestimation at most backward angles becomes smaller (about $15 - 20$\,\%).
However, at the same time the data between $120^\circ$ and $160^\circ$ are
overestimated by about $10 - 20$\,\%; this leads to the overall significantly
worse $\chi^2/F$ in this fit.

A relatively poor fit with $\chi^2/F$ may also result from an inappropriate
OMP. Although it is very unlikely that the otherwise successful folding
potential \cite{Mohr13} fails in the particular case of \zniv\ at low
energies, we have repeated the above procedure of fitting the full angular
distribution and truncated angular distributions using WS potentials of volume
type in the real and imaginary part of OMP. Almost exactly the same behavior
was found in this case (see lower parts of Figs.~\ref{fig:e12elast} and
\ref{fig:e16elast}).

From all the above calculations the total reaction cross sections have been
determined using Eq.~(\ref{eq:stot}). Fortunately, the results for \stot\ turn
out to be very stable. At the lower energy we find an average value of
\stot\ = $428 \pm 7$\,mb. The highest (lowest) value of \stot\ = 440\,mb
(420\,mb) is found for the folding potential truncated at $135^\circ$
($115^\circ$). At the higher energy we find an average value of
\stot\ = $911 \pm 13$\,mb. The highest (lowest) value of \stot\ = 928\,mb
(895\,mb) is found for the folding potential truncated at $155^\circ$
(WS potential truncated at $135^\circ$). The result of the PSF
(\stot\ = 898\,mb) is relatively close to lowest result, and there seems to be
a small systematic deviation between the folding potential fits (average
\stot\ = $925 \pm 3$\,mb) and the WS potential fits (average \stot\ = $902 \pm
5$\,mb). Therefore, we recommend \stot\ = $905 \pm 18$\,mb (with a slightly
increased 2\,\% uncertainty) at the higher energy.

\begin{figure}[h]
\includegraphics[width=0.95\columnwidth,clip=]{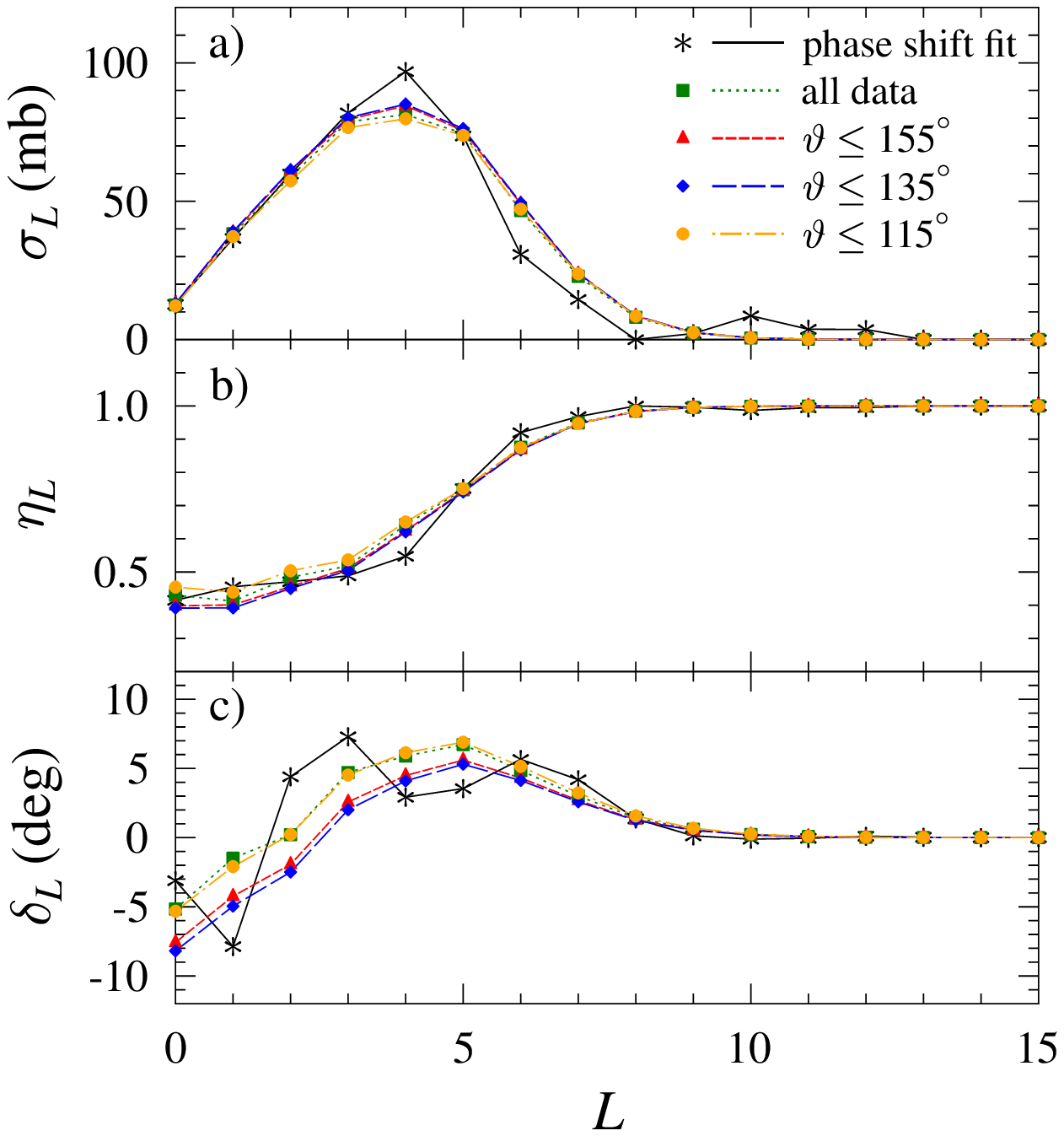}
\caption{
\label{fig:e12phase}
(Color online)
Phase shifts $\delta_L$, reflexion coefficients $\eta_L$, and contribution
$\sigma_L$ of the $L^{\rm{th}}$ partial wave to \stot\ in Eq.~(\ref{eq:stot})
at $E_{\rm{lab}} = 12.05$\,MeV for the OMP fits using a folding potential in
the real part and a surface WS potential in the imaginary part (from bottom to
top). Further discussion see text.
}
\end{figure}

\begin{figure}[h]
\includegraphics[width=0.95\columnwidth,clip=]{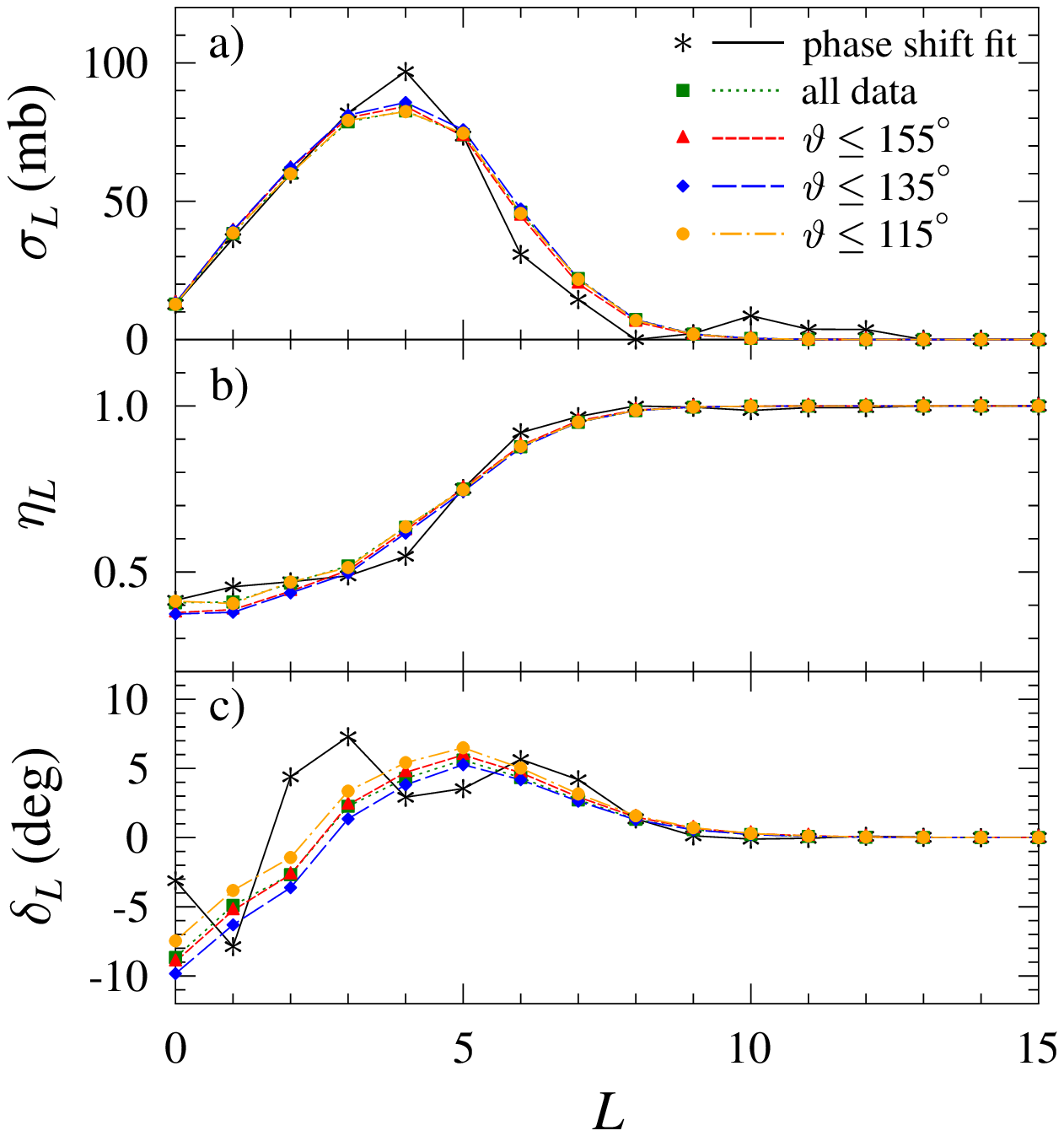}
\caption{
\label{fig:e12phase_ws}
(Color online)
Same as Fig.~\ref{fig:e12phase}, but for the OMP fits with a volume WS
potential in the real and imaginary part of the nuclear potential.
Further discussion see text.
}
\end{figure}

\begin{figure}[h]
\includegraphics[width=0.95\columnwidth,clip=]{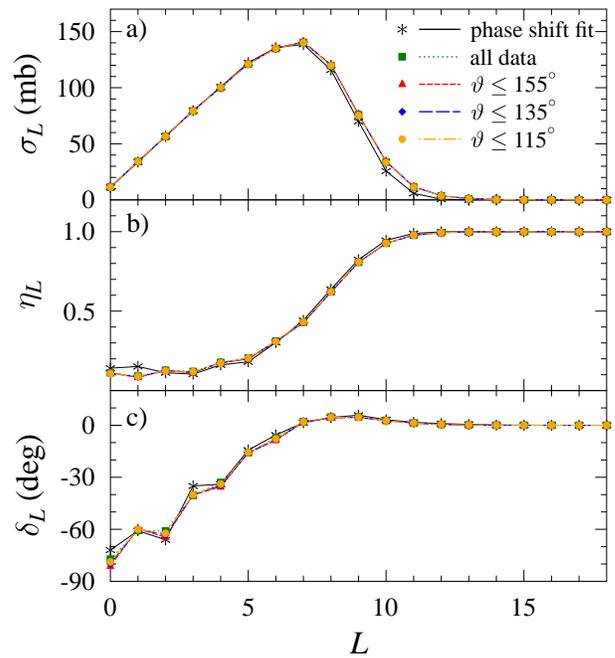}
\caption{
\label{fig:e16phase}
(Color online)
Same as Fig.~\ref{fig:e12phase}, but at $E_{\rm{lab}} = 16.12$\,MeV.
Further discussion see text.
}
\end{figure}

\begin{figure}[h]
\includegraphics[width=0.95\columnwidth,clip=]{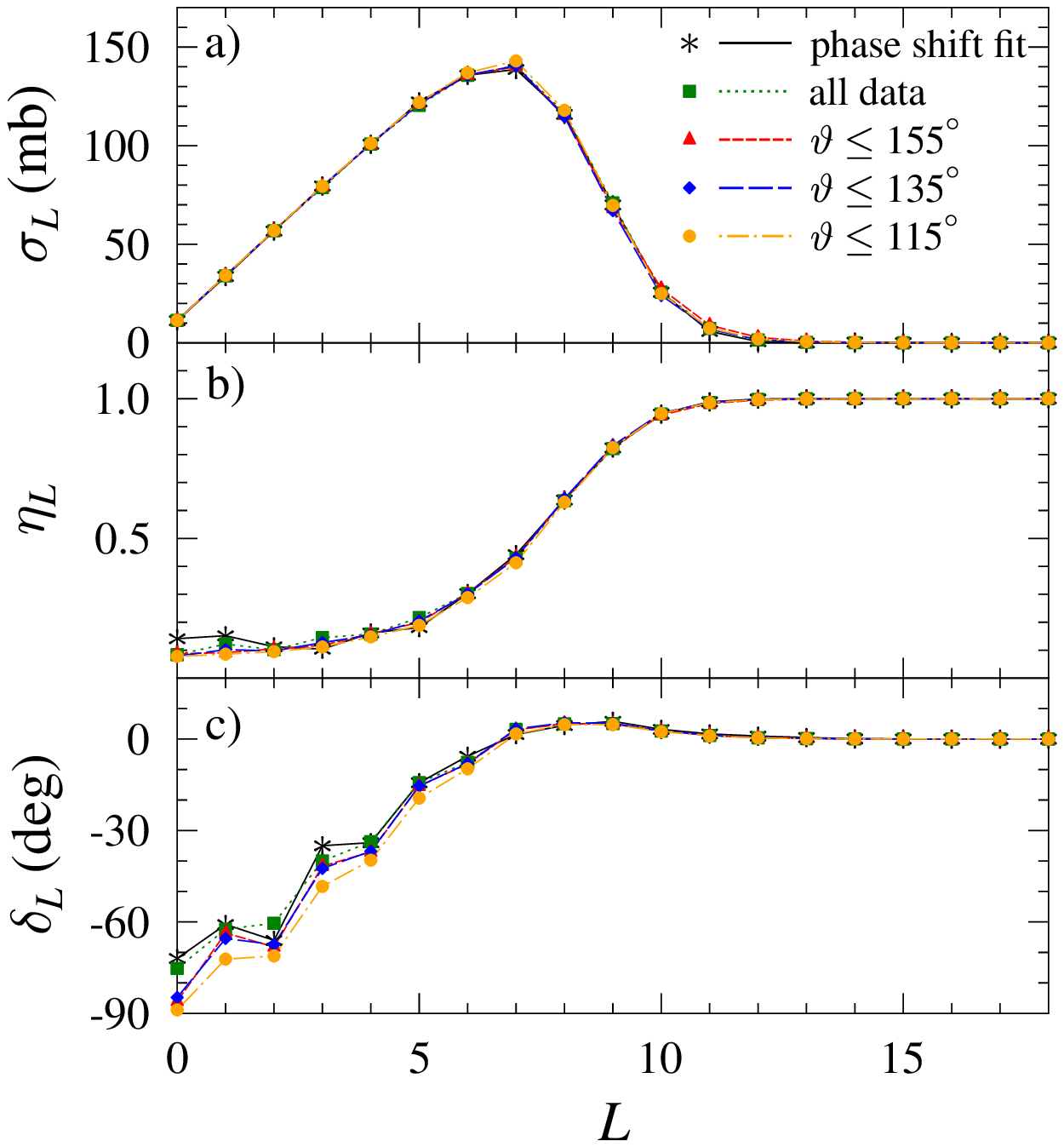}
\caption{
\label{fig:e16phase_ws}
(Color online)
Same as Fig.~\ref{fig:e16phase}, but for the OMP fits with a volume WS
potential in the real and imaginary part of the nuclear potential.
Further discussion see text.
}
\end{figure}

For a better understanding of the differences between the PSFs and the OMP
fits we show the reflexion coefficients $\eta_L$ and the real phase shifts
$\delta_L$ at both energies in Figs.~\ref{fig:e12phase}--\ref{fig:e16phase_ws}.

At the lower energy of 12.1\,MeV, there is an obvious discrepancy between the
PSF and the OMP fits. The OMP fits lead to a relatively smooth variation of
the phase shifts $\delta_L$ with the angular momentum number $L$. Contrary to
the OMP fits, the PSF shows stronger variations in $\eta_L$ and $\delta_L$
which cannot be reproduced by a typical \al -nucleus potential. It is
interesting to note that although the underlying reflexion coefficients
$\eta_L$ are not identical, the resulting total reaction cross section
\stot\ is almost the same for all fits (see above).

At the higher energy of 16.1\,MeV the discrepancies between the $\eta_L$ and
$\delta_L$ from the PSF and from the
OMP fits are smaller. This is not surprising as the backward rise at the higher
energy is not as pronounced as at 12.1\,MeV. The small systematic discrepancy
between the OMP fits using either a folding potential or a WS potential is
mainly related to tiny differences in $\eta_L$ for $8 \le L \le 11$.

\subsection{Reduced reaction cross section \sred }
\label{sec:sred}
\begin{figure}[htb]
\includegraphics[width=\columnwidth,clip=]{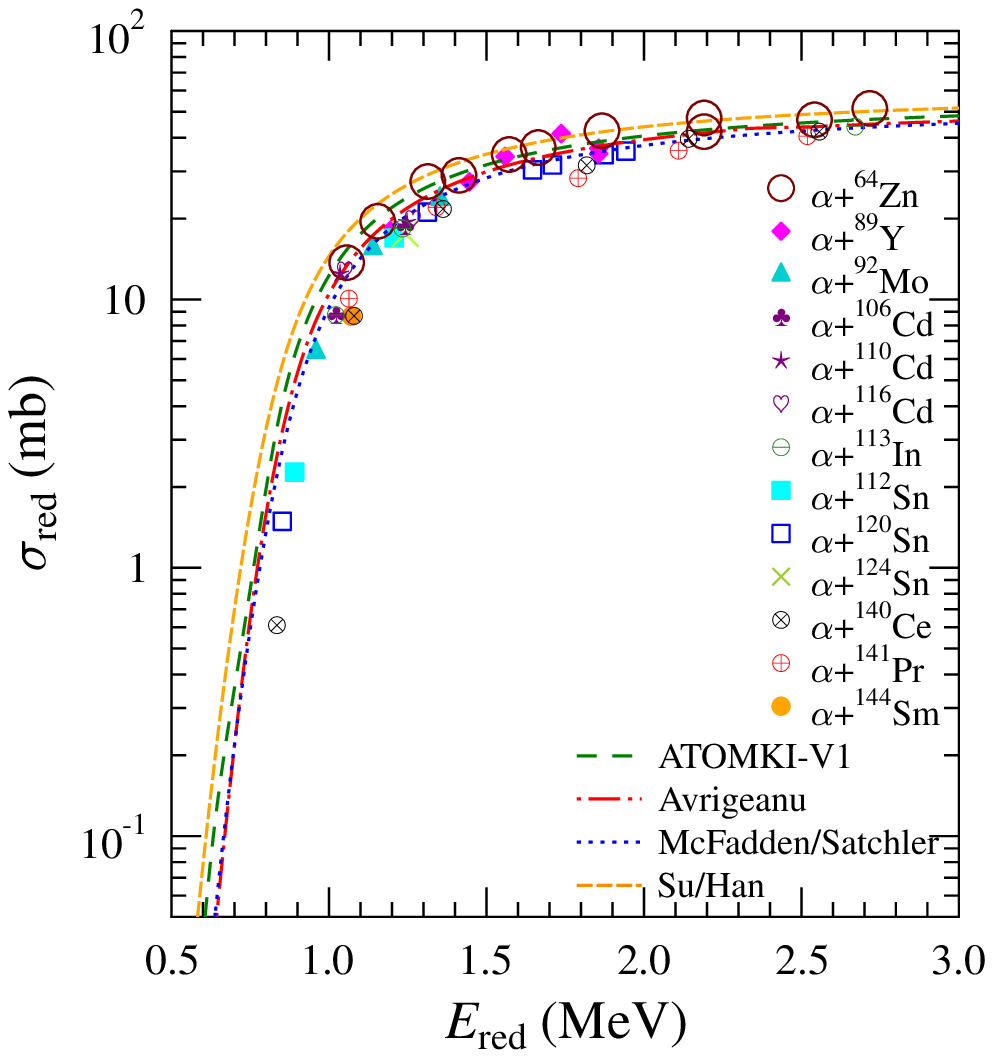}
\caption{
\label{fig:sigred}
(Color online)
Reduced reaction cross sections $\sigma_{\rm{red}}$ versus reduced energy
$E_{\rm{red}}$ for various targets in a broad energy range. Except for \zniv ,
the data are taken from \cite{Mohr13,Mohr11,Mohr13b,Kis13}.
}
\end{figure}

Total reaction cross sections \stot\ of \al -induced reactions for many target
nuclei and in a broad energy range follow a systematic behavior which becomes
visible in a plot of so-called reduced cross sections \sred\ versus reduced
energy \ered\ as suggested in \cite{Gom05}.
\begin{eqnarray}
E_{\rm{red}} & = & \frac{\bigl(A_P^{1/3}+A_T^{1/3}\bigr) E_{\rm{c.m.}}}{Z_P Z_T} \\
\sigma_{\rm{red}} & = & \frac{\sigma_{\rm{reac}}}{\bigl(A_P^{1/3}+A_T^{1/3}\bigr)^2}
\label{eq:red}
\end{eqnarray}
The result is shown in Fig.~\ref{fig:sigred}. Contrary to the common trend for
all nuclei with masses above $A \approx 90$, the data for \zniv\ are slightly
higher than the general trend at all energies under study. Very recently, the
analysis of reaction data for lighter targets ($^{23}$Na \cite{Alm14} and
$^{33}$S \cite{Bow13}) has shown that \sred\ for these light nuclei is
dramatically higher than the general trend for heavy nuclei
\cite{Mohr14}. However, the dramatically increased reduced cross sections for
$^{23}$Na from the $^{23}$Na\rap $^{26}$Mg data of \cite{Alm14} were not
confirmed by later experiments \cite{How15PRL,Tom15PRL} and turned out to be
an experimental error \cite{Alm15}.
Fig.~\ref{fig:sigred} shows also the predictions from four \al -nucleus
potentials \cite{McF66,Avr10,Mohr13,Su15}. These results will be discussed
later.

For completeness it should be noted that there is an approximate relation
between reduced energies and the Gamow window \cite{Mohr16}: $E_{{\rm{red}},0}
\approx 0.284\,{\rm{MeV}} \times T_9^{2/3}$. Consequently, the astrophysically
relevant range for the reduced energy \ered\ is located below the shown range
of Fig.~\ref{fig:sigred} which was chosen from the availability of experimental
scattering data.

\subsection{ALAS for \zniv ?}
\label{sec:alas}
Anomalous large-angle elastic scattering (ALAS) has been discussed in
literature already many years ago (e.g., \cite{Lan82,Mic95}). However, there
is no strict definition for ALAS. The phenomenon was first discussed in
connection with
$^{16}$O\raa $^{16}$O and $^{40}$Ca\raa $^{40}$Ca elastic scattering. For
these reactions it was noticed later that the so-called anomalous cross
sections are related to weak absorption of the doubly-magic target nuclei, and
it is possible to describe the angular distributions within the OM
\cite{Lan82,Mic95,Abe93,Atz96,Hir13,Abd03}. Contrary to these findings, it
turned out that the reproduction of the backward angular range in
e.g.\ $^6$Li\raa $^6$Li or $^{20}$Ne\raa $^{20}$Ne remains extremely difficult,
and various explanations for the backward rise have been suggested: inelastic
coupling to low-lying excited states, compound-elastic contributions, elastic
\al\ transfer, angular-momentum-dependent absorption
(e.g., \cite{Bac72,Sam92,Yan11,Fri71}).

As we have seen above, fortunately the influence of the backward rise on the
derived total cross section \stot\ remains small. Therefore, a complete
theoretical analysis of the backward rise remains beyond the scope of the
present paper. Nevertheless, two of the above effects will be analyzed in more
detail. Firstly, inelastic scattering may contribute significantly to the total
reaction cross section \stot . For a quantitative analysis we have measured
several angular distributions (See Sect.~\ref{sec:inelast}). Secondly,
compound-elastic scattering may contribute to the elastic scattering angular
distribution but by definition it is not included in the OM analysis (see
Sect.~\ref{sec:compound}).

\subsection{Analysis of inelastic scattering}
\label{sec:inelast}
One focus of the present study is the comparison between the total reaction
cross section \stot\ from elastic scattering in Eq.~(\ref{eq:stot}) and the
sum over all open non-elastic channels. This will be discussed in further
detail in Sec.~\ref{sec:comp}. Besides the real reaction channels like \ran ,
\rap , and \rag , inelastic \raapr\ may contribute to this sum. In our
previous study \cite{Gyu12} we have estimated the \raapr\ cross section from
coupled-channel calculations and from Coulomb excitation. Now we are able to
provide experimental constraints for the \raapr\ cross section for the
low-lying states.

A precise experimental determination of the total \raapr\ cross section is
very difficult for at least two reasons. The total inelastic \raapr\ cross
section is composed of contributions to all excited states in the target
nucleus \zniv\ with excitation energies $E_x$ below $E_{\rm{c.m.}}$ of the
scattering experiment. In practice, the spectra in
Fig.~\ref{fig:scatteringspectra} allow a determination of the \raapr\ cross
section only for the lowest excited states in \zniv , and in particular at
backward angles a significant yield appears about 5\,MeV below the elastic
peak. Besides inelastic scattering from \zniv , this yield may also come from
reaction products of \al -induced reaction on the \zniv\ target and all target
contaminations because the detectors do not allow the identification of the
ejectiles. Furthermore, the measurement of each angular distribution is
complicated because inelastic peaks may overlap with elastic scattering from
lighter nuclei in the target (e.g., in the carbon backing). In addition, at
forward angles the \raapr\ cross section is much smaller than the elastic
cross section which approaches the Rutherford cross section and thus increases
dramatically to small scattering angles with with $1/\sin^4{(\vartheta/2)}$.

\subsubsection{Inelastic scattering to low-lying excited states}
\label{sec:low}
The first excited states in the level scheme of \zniv\ consist of a $2^+$
state at 992\,keV and a triplet of states ($0^+$, $2^+$, $4^+$) with almost
twice the excitation energy of the first $2^+$ state, i.e.\ a typical
vibrational behavior. Thus, these inelastic angular distributions were analyzed
within the anharmonic vibrator model which is implemented in the widely used
coupled-channels ECIS code \cite{ECIS}.

Experimental angular distributions were measured for the first $2^+$ state at
992\,keV and the $4^+$ state at 2307\,keV. The experimental resolution was not
sufficient to separate the second $2^+$ state at 1799\,keV and the $0^+$ state
at 1910\,keV; only the sum of both states could be determined. For
experimental details, see Sec.~\ref{sec:exp_scat}.

\begin{figure}[htb]
\includegraphics[width=0.95\columnwidth,clip=]{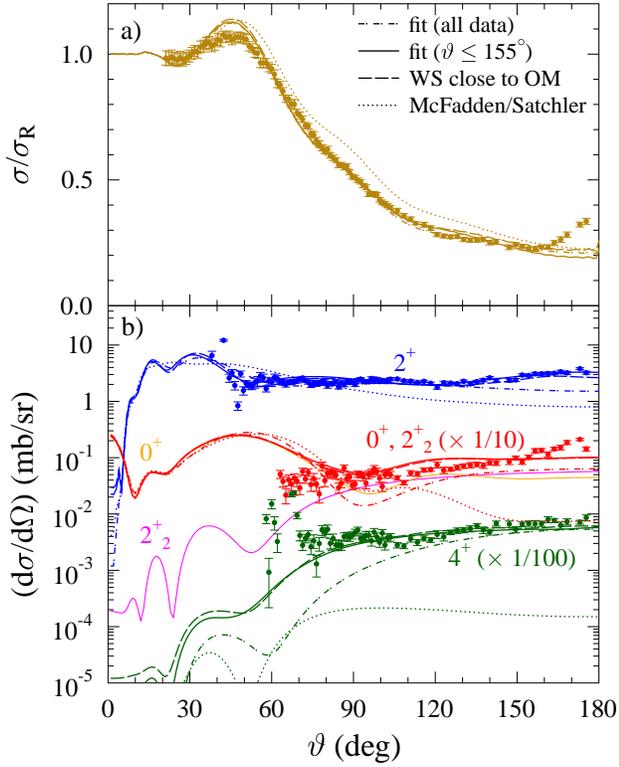}
\caption{
\label{fig:e12inelast}
(Color online)
Inelastic \zniv \raapr \zniv\ scattering cross sections at the energy
12.1\,MeV. The upper part (a) shows the elastic angular distribution
(normalized to Rutherford). The lower part (b) shows the inelastic angular
distributions for the first $2^+$ state (blue), the sum of $0^+$ and $2^+_2$
(red), and the $4^+$ state (green). In addition, for the best-fit calculation
(full lines) a decomposition into the $0^+$ (orange) and $2^+_2$ states
(magenta) is shown. For explanation of the various calculations, see text.
}
\end{figure}

\begin{figure}[htb]
\includegraphics[width=0.95\columnwidth,clip=]{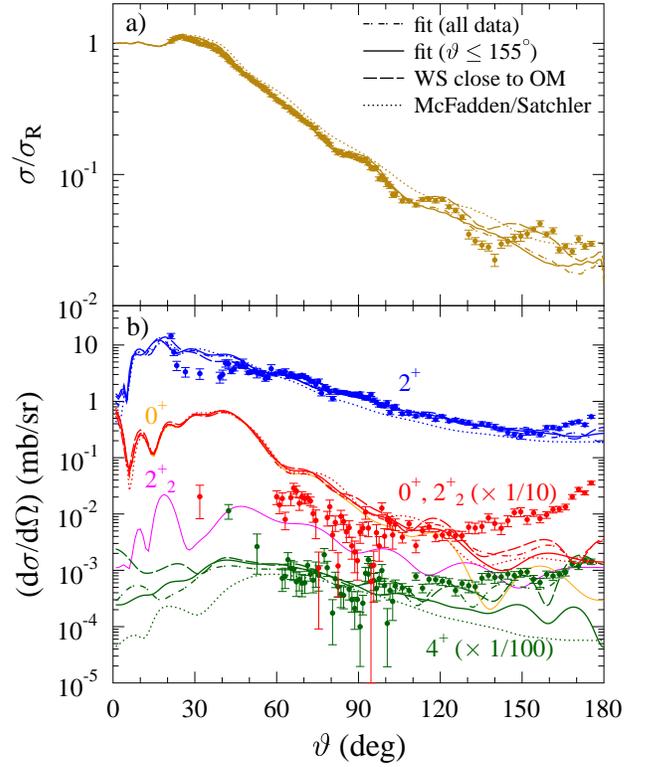}
\caption{
\label{fig:e16inelast}
(Color online)
Inelastic \zniv \raapr \zniv\ scattering cross section at the energy
16.1\,MeV. For explanations, see Fig.~\ref{fig:e12inelast}.
}
\end{figure}

It has been difficult to fit the elastic scattering angular
distributions, see Sec.~\ref{sec:elast_new}. Obviously, these problems appear
also when simultaneous fits 
are made to elastic and inelastic angular distributions. Several fits with
different potentials and a varying number of adjustable parameters have been
made. These fits show significant differences in the reproduction of the
angular distributions, but fortunately the angle-integrated inelastic cross
sections are quite stable.  The results are shown in Fig.~\ref{fig:e12inelast}
for the 12.1\,MeV data and in Fig.~\ref{fig:e16inelast} for the 16.1\,MeV
data.

As a first approximation, the widely used potential of McFadden and Satchler
\cite{McF66} was applied in combination with an adjustment of the couplings to
the inelastic states (dotted lines in Figs.~\ref{fig:e12inelast} and
\ref{fig:e16inelast}). The reproduction of the elastic angular distributions
is reasonable but not perfect. Such a behavior is expected because the
potential parameters were not re-adjusted. The angular distribution of the
$2^+_1$ state is reasonably well described, but the two-phonon states and in
particular the $4^+$ state cannot be reproduced.

In a next step, the Woods-Saxon potential from the optical model fits
(restricted to scattering angles below $155^\circ$) was used (dashed lines,
label ``WS close to OM''); see also Sec.~\ref{sec:elast_new}. Again, the
coupling to the inelastic states was fitted, and a minor readjustment was
allowed for the depths of the real and imaginary parts of the Woods-Saxon
potential. Of course, from the fitting procedure the agreement for the elastic
angular distribution improves. But simultaneously also the description of the
inelastic angular distributions improves.

In a third calculation, additionally all parameters of the Woods-Saxon
potentials (real and imaginary depths, radii, and diffusenesses) were adjusted
simultaneously (full lines, label ``fit ($\vartheta \le
155^\circ$)''). Although a smaller $\chi^2$ is obtained, the visible changes
in the angular distributions remain relatively small.

The final calculation repeats the third calculation, but includes all
experimental data, i.e.\ including the backward rise of the elastic angular
distributions beyond $\vartheta \approx 155^\circ$. The results (dash-dotted
lines, label ``fit (all data)'') show significantly worse agreement in the
backward angular region of the inelastic angular distributions. This is
related to a wide variation of the WS parameters, similar to the problems
found in the OM study in Sec.~\ref{sec:elast_new}. Hence, the backward rise of
the elastic cross sections cannot be explained by the inelastic coupling to
low-lying excited states.

Interestingly, despite the relatively wide changes of the inelastic angular
distributions, the angle-integrated inelastic cross sections remain relatively
stable. E.g., at 12.1\,MeV for the dominating first $2^+$ state a cross
section of 27.3\,mb is obtained from the McFadden/Satchler potential, the
Woods-Saxon potential from optical model fit gives 32.9\,mb, and the fits to
data up to $155^\circ$ (all data) result in 34.3\,mb (30.1\,mb). Excluding the
McFadden/Satchler result (without any adjustment of the potential to the
experimental data), we adopt a semi-experimental angle-integrated cross
section of $33 \pm 3$\,mb in this case. Similar results are found for all
angle-integrated inelastic cross sections at both experimental energies. The
results are listed in Table \ref{tab:inelast}. For each level, the given
uncertainties are estimated from the variations of the different
fits. However, it should be kept in mind that the experimental inelastic
angular distributions do not cover the full angular range and thus cannot
fully constrain the fits; this holds in particular for the significant
contribution of the $0^+_2$ state at forward angles. Therefore, a somewhat
increased uncertainty of about 15\,\% is carefully estimated for the sum over
the experimentally determined inelastic cross sections to low-lying excited
states in Table \ref{tab:inelast}.
\begin{table*}[htb]
\caption{
\label{tab:inelast} 
Angle-integrated inelastic cross sections (in mb).
}
\begin{tabular}
{|cr@{\,$\pm$\,}l|r@{\,$\pm$\,}lr@{\,$\pm$\,}lr@{\,$\pm$\,}lr@{\,$\pm$\,}lr@{\,$\pm$\,}l|r@{\,$\pm$\,}l|}
\hline
E$_\alpha$ & \multicolumn{2}{c|}{E$^{\rm eff.}_{\rm c.m.}$} 
& \multicolumn{2}{c}{$2^+_1$ (992\,keV)} 
& \multicolumn{2}{c}{$2^+_2$ (1799\,keV)}
& \multicolumn{2}{c}{$0^+_2$ (1910\,keV)}
& \multicolumn{2}{c}{$4^+_1$ (2307\,keV)}
& \multicolumn{2}{c|}{$\Sigma_{\rm{low}}^{\rm{exp}}$}
& \multicolumn{2}{c|}{$\Sigma_{\rm{high}}^{\rm{calc}}$} \\
\hline
12.05 & 11.29 & 0.09 & 33 & 3 & 3.1 & 0.5 & 10.2 & 0.5 & 3.2 & 0.5 
& 49.5 & 7.5\footnote{discussion of uncertainty: see text}
& 31.5 & 6.3\footnote{estimated uncertainty of 20\,\% (see text)} \\
16.12 & 15.17 & 0.08 & 28 & 1 & 0.6 & 0.2 & 13.6 & 0.5 & 0.9 & 0.1 
& 43.1 & 6.5\footnotemark[1] 
& 157.9 & 30.6\footnotemark[2]  \\
\hline
\end{tabular}
\end{table*}

\subsubsection{Inelastic scattering to higher-lying excited states}
\label{sec:high}
The inelastic \raapr\ cross sections to higher-lying states above the ($0^+$,
$2^+$, $4^+$)-triplet were estimated using the combination of direct and
compound contributions as implemented in the widely used nuclear reaction code
TALYS (version 1.8). Contrary to the
first excited states with their dominating direct contributions, inelastic
scattering to higher-lying states is dominated by compound contributions. In
these TALYS calculations 30 low-lying levels below $E^\ast = 3.5$\,MeV in
\zniv\ were taken into account explicitly; for $E^\ast > 3.5$\,MeV a continuum
contribution is estimated using a theoretical level density.

The first 4 excited states were already taken into account in the previous
section \ref{sec:low}. Thus, the summed inelastic cross section to
higher-lying states $\sum_{\rm{high}}$ is estimated from the calculated total
inelastic cross section in TALYS minus the calculcated inelastic cross
sections to the first 4 excited states. Fortunately, even a broad variation
of the TALYS parameters (mainly a variation of the \al -nucleus potential)
leads to relatively stable values for the inelastic cross sections. At the
lower energy 
of 12.1\,MeV, $\sum_{\rm{high}}$ varies between 26.6\,mb and 34.3\,mb with an
average value of $31.5 \pm 2.8$\,mb. Because of the missing experimental
constraint, we finally assign a larger 20\,\% uncertainty to this value. At
the higher energy of 16.1\,MeV $\sum_{\rm{high}}$ shows significantly larger
values between 139\,mb and 168\,mb with an average of $157.9 \pm
11.1$\,mb. Again, we finally assign a 20\,\% uncertainty (see Table
\ref{tab:inelast}). Indeed, this choice of the uncertainty for
$\sum_{\rm{high}}$ is somewhat arbitrary. But we think that increasing the
TALYS uncertainties by a factor of about two should provide a careful
estimate of the real uncertainty of $\sum_{\rm{high}}$.

\subsection{Compound-elastic contributions to low-lying states}
\label{sec:compound}
In general, the compound mechanism may also contribute to the elastic angular
distribution and to the inelastic angular distributions of the low-lying
states. The angle-integrated compound-elastic contribution is small with about
$5-7$\,mb at 12.1\,MeV and around 0.5\,mb at 16.1\,MeV. Angle-integrated
compound-inelastic cross sections to the low-lying states in
Sec.~\ref{sec:low} are of the order of a few mb at 12.1\,MeV and below 1\,mb
at 16.1\,MeV (calculated by TALYS, again mainly varying the \al -nucleus
potential). Thus, the compound contributions may slightly affect the elastic 
angular distribution in particular at backward angles where the direct cross
section is small, and may also contribute to the unexpected elastic cross
sections at backward angles (see Sec.~\ref{sec:alas}). The inelastic angular
distributions will also be somewhat affected by the compound contributions;
however, as long as a reasonable description of the experimental angular
distributions is achieved within the direct coupled-channels model, the
angle-integrated inelastic cross section should be well-defined by these
calculations in the coupled-channels approach (as done in Sec.~\ref{sec:low}).

\section{Comparison of total cross sections from elastic scattering to
  reaction cross sections}
\label{sec:comp}
The total non-elastic cross section \stot\ is given by the sum over all open
channels. For \zniv\ at the energies under study this means:
\begin{eqnarray}
\sigma_{\rm{reac}} & = &
  \sigma(\alpha,\alpha') +
  \nonumber \\
& &
  \sigma(\alpha,\gamma) +
  \sigma(\alpha,n) +
  \sigma(\alpha,p) +   
  \nonumber \\
& &
  \sigma(\alpha,\alpha p) +
  \sigma(\alpha,\alpha n) +
  \sigma(\alpha,2\alpha) +
  \sigma(\alpha,2p)
  \nonumber \\
& &
  \sigma(\alpha,d) +
  \sigma(\alpha,t) +
  \sigma(\alpha,{\rm{{^3}He}})
\label{eq:sum}
\end{eqnarray}
\stot\ in the above Eq.~(\ref{eq:sum}) can be derived from
Eq.~(\ref{eq:stot}), i.e.\ from the angular
distribution of elastic scattering (see Sec.~\ref{sec:elast_new} and Table
\ref{tab:pot}). The determination of the sum on the right-hand-side of
Eq.~(\ref{eq:sum}) will be discussed in detail below.

The identity of \stot\ from elastic scattering in Eq.~(\ref{eq:stot}) and from
the sum over non-elastic channels in Eq.~(\ref{eq:sum}) follows directly from
basic quantum-mechanics. In our previous work \cite{Gyu12}
we have calculated the ratio $r$
between the result from elastic scattering and the sum over the non-elastic
channels, and $r = 1$ was found within the experimental uncertainties. As
there is no reason to question this theoretically expected ratio of $r 
\equiv 1$, we can also reverse the arguments: fixing \stot\ from
Eq.~(\ref{eq:stot}) allows to obtain experimental constraints
for the cross sections of unobserved non-elastic channels in the sum on the 
right-hand-side of Eq.~(\ref{eq:sum}). This will become important in
particular for the compound-inelastic \raapr\ cross section at the higher
energy of 16.1\,MeV.

At the lower energy of 12.1\,MeV, all channels with two outgoing particles in
the third line of Eq.~(\ref{eq:sum}) were neglected because these channels are
either closed or have negligible cross sections. The same holds for the last
line in Eq.~(\ref{eq:sum}). The remaining contributions can be taken almost
completely from experiment. The cross sections of the \rag , \ran , and
\rap\ reactions in the second line of Eq.~(\ref{eq:sum}) are taken from Table
\ref{tab:actres}; their sum amounts to $366 \pm 40$\,mb. The \raapr\ inelastic 
scattering cross section is taken from experiment for the low-lying excited
states with $49.5 \pm 7.5$\,mb and from theoretical estimates for the
higher-lying states with $31.5 \pm 6.3$\,mb (see Table
\ref{tab:inelast}). Summing up all these values leads to a total non-elastic
cross section of $447 \pm 41$\,mb. Within the uncertainties, this result is in
excellent agreement with the value of $428 \pm 7$\,mb derived from elastic
scattering (see Table \ref{tab:pot}). As expected, the ratio $r$ between the
result from elastic scattering and the sum over the contributing channels
results in $r = 0.957 \pm 0.090$, i.e.\ it is identical to unity within the
uncertainties. Compared to our previous study \cite{Gyu12}, the uncertainty of
the ratio $r$ could be reduced by a factor of two. This reduction is based on
improved scattering data at the same energy of the reaction cross
sections. Note that the sum in Eq.~(\ref{eq:sum}) is based on experimental
data with the only exception of inelastic scattering to higher-lying states in
\zniv\ which contributes only with 31.5\,mb (or 7\,\%) to the sum of 447\,mb.

At the higher energy of 16.1\,MeV the situation is somewhat more complicated
because more channels are open. Nevertheless, because of strongly negative
$Q$-values, the cross sections of the ($\alpha$,$d$) and ($\alpha$,$^3$He)
reactions remain negligibly small, and the ($\alpha$,$t$) channel is still
closed; thus, the forth line in Eq.~(\ref{eq:sum}) still can be neglected. The
same holds for the reactions with two outgoing particles in the third line of
Eq.~(\ref{eq:sum}) with the exception of the ($\alpha$,$2p$)
reaction. According to TALYS calculations, this reaction contributes with
$17.6 - 18.4$\,mb. Because there is no experimental constraint, we estimate a
contribution of 18\,mb with an uncertainty of 25\,\%.
The other contributions are determined in the same way as for
the lower energy of 12.1\,MeV. The sum of the \rag , \ran , and \rap\ cross
sections amounts to $668 \pm 72$\,mb. The \raapr\ inelastic scattering cross
section is composed of the experimental result for the low-lying excited
states ($43.1 \pm 6.5$\,mb) and of the theoretical estimates for the
higher-lying states ($157.9 \pm 30.6$\,mb). The sum of $887 \pm 79$\,mb again
agrees well with the result of $905 \pm 18$\,mb from elastic scattering,
leading to a ratio of $r = 1.020 \pm 0.093$ between the result from elastic
scattering and the sum over the reaction channels.

As pointed out above, the expected ratio $r \equiv 1.0$ allows to constrain
the cross sections of unobserved non-elastic channels. For the data at the
higher energy of 16.1\,MeV this leads to the conclusion that the significant
contribution of compound-inelastic scattering to higher-lying excites states
of $158 \pm 31$ mb is indeed confirmed experimentally by the present
data. Starting from the experimental result from elastic scattering in
Eq.~(\ref{eq:stot}) and subtracting the experimentally determined cross
sections, i.e.\ the inelastic scattering to low-lying states and the reaction
channels in the second line of Eq.~(\ref{eq:sum}), this leads to a remaining
cross section of $194 \pm 75$\,mb which has to be distributed among the
remaining open channels which are mainly inelastic scattering to higher-lying
states ($\approx 158$\,mb from TALYS) and to a minor degree the
($\alpha$,$2p$) reaction ($\approx 18$\,mb from TALYS).

\section{Predictions from global \al -nucleus potentials}
\label{sec:pred}
Finally, the experimental results for the total reaction cross sections
\stot\ and for the \rag , \ran , and \rap\ cross sections are compared to
predictions from global \al -nucleus potentials. Interestingly, it turns out
that the widely used \al -nucleus potentials predict very similar
total cross sections. The predictions of Watanabe (TALYS default)
\cite{Wat58}, McFadden and Satchler \cite{McF66}, Demetriou {\it et
  al.}\ \cite{Dem02}, Avrigeanu {\it et al.}\ \cite{Avr14}, Su and Han
\cite{Su15}, and from the ATOMKI-V1 potential \cite{Mohr13} are listed in
Table \ref{tab:predict}. The new potential by Su and Han \cite{Su15}
overestimates the total reaction cross section in particular at low energies;
at 5\,MeV the predicted \stot\ from the potential by Su and Han is about
11\,$\mu$b whereas the predictions from the other potentials
\cite{Wat58,McF66,Dem02,Avr14,Mohr13} vary between 0.3\,$\mu$b and 1.6\,$\mu$b
with a mean value of about 0.8\,$\mu$b.
(For completeness it may be noted here that the
present TALYS 1.8 version uses still the Watanabe potential as default,
although the TALYS manual states that this has changed to Avrigeanu {\it et
  al.}~\cite{Avr14}.)
\begin{table}[htb]
\caption{
\label{tab:predict}
Predictions of \stot\ from various global \al\ -nucleus potentials, compared
to the results from elastic scattering. All cross sections are given in mb.
}
\begin{tabular}{ccl}
\hline
\multicolumn{1}{c}{E$_\alpha = 12.05$\,MeV} &
\multicolumn{1}{c}{E$_\alpha = 16.12$\,MeV} &
Ref. \\
\hline
$428 \pm 7$ & $905 \pm 18$ & experiment (this work) \\
408 & ~833 & Watanabe \cite{Wat58} \\
376 & ~809 & McFadden/Satchler \cite{McF66} \\
427 & ~868 & Demetriou {\it et al.}, V1 \cite{Dem02} \\
405 & ~837 & Demetriou {\it et al.}, V2 \cite{Dem02} \\
455 & ~845 & Demetriou {\it et al.}, V3 \cite{Dem02} \\
415 & ~857 & Avrigeanu {\it et al.}\ \cite{Avr14} \\
475 & ~915 & ATOMKI-V1 \cite{Mohr13} \\
552 & 1010 & Su and Han \cite{Su15} \\
\hline
\end{tabular}
\end{table}

\begin{figure}[htb]
\includegraphics[width=0.95\columnwidth,clip=]{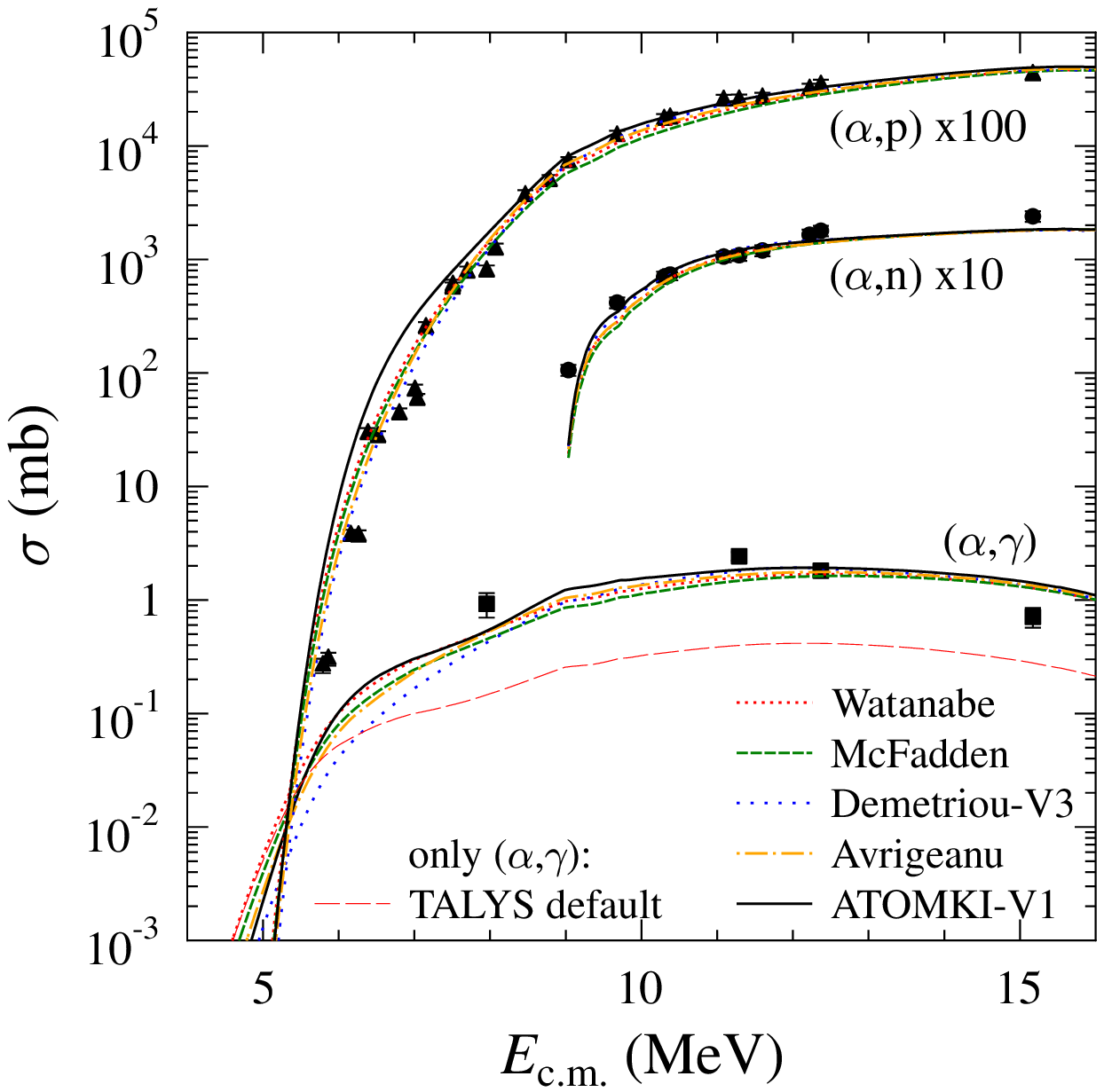}
\caption{
\label{fig:reac}
(Color online)
Cross sections of the \zniv \rag \geviii , \zniv \ran \gevii , and \zniv \rap
\gavii\ reactions. The experimental data are taken from our previous work
\cite{Gyu12} except the new results at the highest energy $E_\alpha =
16.1$\,MeV. The calculations are based on different global \al -nucleus
optical potentials \cite{Wat58,McF66,Dem02,Avr14,Mohr13}. For better
readability, two versions of \cite{Dem02} have been omitted. The differences
between the predictions from various \al -nucleus potentials are relatively
small. TALYS default
parameters have been used in general except for the $\gamma$-ray strength
function; the default $\gamma$-ray strength underestimates the \rag\ cross
section (thin red long-dashed line). Further discussion see text.
}
\end{figure}

The total reaction cross section \stot\ is mainly composed of the dominating
\rap\ and \ran\ channels. The branching between these two channels is
sensitive to the chosen nucleon potential whereas the total reaction cross
section \stot\ is sensitive only to the \al -nucleus potential. The TALYS
default option for the nucleon potential \cite{Kon03} works very well here and
was not changed in this work. Excellent agreement for the \ran\ and
\rap\ cross sections is found, see Fig.~\ref{fig:reac}.

The cross section of the \zniv \rag \geviii\ reaction is sensitive to the \al
-nucleus potential and to the $\gamma$-ray strength function. Here the best
result is obtained using the Hartree-Fock BCS $\gamma$-ray strength from
\cite{Cap09}. The TALYS default option using generalized Lorentzian
$\gamma$-ray strength from \cite{Kop90} significantly underestimates the \zniv
\rag \geviii\ cross section (see Fig.~\ref{fig:reac}, thin red long-dashed
line).

\section{Summary and conclusions}
\label{sec:summ}
Elastic and inelastic \zniv \raa \zniv\ scattering was measured at the
energies of 12.1\,MeV and 16.1\,MeV. At the same energies the cross sections
of the \rag , \ran , and \rap\ reactions were determined using the activation
technique. The experimental angular distributions of elastic scattering cover
the full angular range and thus allow for a precise determination of the total
non-elastic reaction cross section \stot\ with uncertainties of about
$2-3$\,\%. A perfect description of the elastic angular distributions could
only be achieved using phase shift fits. The surprising rise of the elastic
cross sections at very backward angles may be considered as so-called ALAS and
could not be fully explained. Fortunately, the behavior of the angular
distibutions at these very backward angles practically does not affect the
determination of the total cross sections \stot .

The total reaction cross sections follow a general smooth trend when presented
as so-called reduced cross sections \sred\ versus reduced energies \ered . The
data for \zniv\ lie in between the common behavior for heavy target nuclei
with $A \gtrsim 90$ \cite{Mohr13} and slightly increased values for lighter
target nuclei with $A \lesssim 50$ \cite{Mohr15}.

The total non-elastic reaction cross section was also determined from the sum
over the cross sections of all non-elastic channels (including inelastic
scattering). At the lower energy of 12.1\,MeV excellent agreement between
\stot\ from elastic scattering and from the sum over non-elastic channels was
found; here almost all open channels (including inelastic scattering to
low-lying states in \zniv ) could be determined experimentally. At the higher
energy of 16.1\,MeV we find again excellent agreement for \stot\ from the two
approaches. However, now a significant contribution of inelastic scattering to
higher-lying states in \zniv\ is required which is indeed predicted in the
statistical model. In turn, this may be considered as an experimental
verification of these statistical model predictions. Compared to our previous
work \cite{Gyu12}, the experimental uncertainties in the comparison of
\stot\ from the two approaches could be reduced significantly by about a
factor of two.

Usually, the cross sections of \al -induced reactions in the statistical model
depend sensitively on the chosen \al -nucleus potential. At 12.1\,MeV and
16.1\,MeV, the recent global \al -nucleus potentials predict very similar total
reaction cross sections \stot , and with the exception of the latest potential
by Su and Han \cite{Su15} this behavior surprisingly persists down
to lower energies. Thus, in the case of \zniv\ the total reaction cross
section \stot\ can be described well, and the \rag , \ran , and \rap\ data can
be used to constrain further ingredients of the statistical model. In
particular, it is found that the TALYS default nucleon optical potential
\cite{Kon03} works very well, whereas the \rag\ data can be best described
using the Hartree-Fock BCS $\gamma$-ray strength \cite{Cap09} but the
default generalized Lorentzian \cite{Kop90} significantly underestimates the
experimental data.

\begin{acknowledgments}
This work was supported by 
OTKA (K108459 and K120666),
NTKH/FCT Bilateral cooperation program 6818,
and Technological Research Council of Turkey
(TUBITAK-Grant-109T585). 
A.\ O.\ acknowledges support from ERASMUS scholarship, and
M.\ P.\ T.\ acknowledges support from T\'AMOP 4.2.4.\ A/2-11-1-2012-0001
National Excellence Program.
\end{acknowledgments}

\end{document}